\documentclass[]{article}
\usepackage{hyperref}

\usepackage{amsfonts}
\usepackage{amssymb}
\usepackage{amsmath}
\usepackage{bbm}
\usepackage{tikz}

\newcommand{\rem}[1]{{\bf Remark:}}

\def\QED{{\hspace*{\fill}{\vrule height .5ex width 1ex }\quad} 
    \vskip 0pt plus20pt}
\newcommand{\be}{\begin{equation}}
\newcommand{\ee}{\end{equation}}
\newcommand{\bea}{\begin{eqnarray}}
\newcommand{\eea}{\end{eqnarray}}
\newcommand{\beann}{\begin{eqnarray*}}
\newcommand{\eeann}{\end{eqnarray*}}

\newcommand{\ket}[1]{\vert{#1}\rangle}
\newcommand{\bra}[1]{\langle{#1}\vert}

\newcommand{\unity}{\boldsymbol{1}}

\newcommand{\Hil}{\mathcal{H}}

\newcommand{\C}{\mathbb{C}}
\newcommand{\Z}{\mathbb{Z}}

\title{Counterexamples to Ferromagnetic Ordering of Energy Levels}
\author{Wolfgang Spitzer$^{1}$, Shannon Starr$^{2}$ and Lam Tran$^{3}$\\[5pt]
${}^{1}$ \small FernUniversit\"at Hagen,
Fakult\"at f\"ur Mathematik und Informatik,\\
\small Hagen, Germany
\href{Wolfgang.Spitzer@FernUni-Hagen.de}{Wolfgang.Spitzer@FernUni-Hagen.de} \normalsize
\\[3pt]
${}^2$ \small
University of Rochester, Department of Mathematics,\\
\small Rochester, NY, 14627, USA
\href{sstarr@math.rochester.edu}{sstarr@math.rochester.edu} \normalsize
\\[3pt]
${}^3$ \small Toyota Technological Institute at Chicago, University of Chicago,\\
\small 6045 S. Kenwood Ave. Chicago, Illinois 60637, USA\\
\small \href{lctran@uchicago.edu}{lctran@uchicago.edu} \normalsize
}
\date{August 3, 2011}

\begin{document}
%\onehalfspacing

\markright{Counterexamples to FOEL}

\maketitle

\begin{abstract}

The Heisenberg ferromagnet has symmetry group ${\rm SU}(2)$.
The property known as ferromagnetic ordering of energy levels (FOEL) states that the minimum
energy eigenvalue among eigenvectors with total spin $s$ is monotone decreasing
as a function of $s$.
While this property holds for certain graphs such as open chains, in this note we demonstrate
some counterexamples.
We consider the spin $1/2$ model on rings of length $2n$ for $n=2,3,\dots,8$, and show that the minimum
energy among all spin singlets is less than or equal to the minimum energy
among all spin triplets, which violates FOEL.
This also shows some counterexamples to the ``Aldous ordering'' for the symmetric exclusion process.
We also review some of the literature related to these examples.

\vspace{8pt}
\noindent
{\small \bf Keywords:} Heisenberg model, quantum spin systems, simple exclusion process, ordering of energy levels,
Aldous ordering, spectral gap, Temperley-Lieb algebra, Bethe ansatz.
\vskip .2 cm
\noindent
{\small \bf MCS numbers:} 82B10, 81R05, 81R50.
\end{abstract}

\newpage

\section{Introduction}

We consider the spin-$1/2$ quantum Heisenberg ferromagnet on even length cycles.
For a positive integer $N$, 
the $N$-cycle is the graph $C_{N}$ with vertex set $\{1,\dots,N\}$ and edges
$$
\{1,2\}\, ,\ \{2,3\}\, ,\ \dots\, ,\ \{N-1,N\}\quad \text { and }\quad \{1,N\}\, .
$$
This is the Cayley graph for the additive cyclic group $\Z/(N\Z)$ with the generator set $S = \{1,-1\}$.
We consider even values: $N=2n$.

The Hilbert space for the cycle $C_N$ is
$$
\Hil_{\rm tot}\, =\, \Hil_1 \otimes \cdots \otimes \Hil_{N}\, ,
$$
with single site Hilbert spaces
$$
\Hil_1,\dots,\Hil_{N}\, \cong\, \C^2\, ,
$$
each with a prescribed orthonormal basis $\{\ket{\uparrow}\, ,\ \ket{\downarrow}\}$.
In this basis, the  $\textrm{SU}(2)$ spin operators are $S^{x,y,z} = \frac{1}{2} \sigma^{x,y,z}$ for 
$$
\sigma^{x}\, =\, \begin{bmatrix} 0 & 1 \\ 1 & 0 \end{bmatrix}\, ,\quad
\sigma^{y}\, =\, \begin{bmatrix} 0 & -i \\ +i & 0 \end{bmatrix}\quad \text { and } \quad
\sigma^{z}\, =\, \begin{bmatrix} +1 & 0 \\ 0 & -1 \end{bmatrix}\, .
$$
Denoting the two-by-two identity matrix as $\unity_{\C^2}$, 
the spin operators at the $k$th site are
$$
S^{x,y,z}_k\, =\, \unity_{\C^2}^{\otimes (k-1)}  \otimes S^{x,y,z} \otimes \unity_{\C^2}^{\otimes (n-k)} \quad \text { for } \quad k=1,\dots,N\, .
$$
Then the spin-$1/2$ ferromagnetic Heisenberg Hamiltonian on $C_{N}$ is
$$
H\, =\, h_1 + \dots + h_{N}\, ,
$$
where for $k=1,\dots,N-1$,
$$
h_k\, =\, (1/4) \unity - {\bf S}_k \cdot {\bf S}_{k+1}\, ,
$$
and 
$$
h_{N}\, =\, (1/4) \unity - {\bf S}_n \cdot {\bf S}_1\, .
$$
We use the notation $\unity$ for the identity operator on $\Hil(C_{N})$, and the usual spin-matrix dot-product is
$$
{\bf S}_k \cdot {\bf S}_{\ell}\, =\, S_k^x S_{\ell}^x + S_k^y S_{\ell}^y + S_k^z S_{\ell}^z\, .
$$
We have shifted the Hamiltonian to have ground state energy equal to zero.

The Hamiltonian operator has the symmetry group $\textrm{SU}(2)$.
It commutes with the total $S^x$, $S^y$ and $S^z$ operators
$$
S^{x,y,z}_{\rm tot}\, =\, S_1^{x,y,z} + \dots + S_{N}^{x,y,z}\, .
$$
It also commutes with the total spin operator, which is sometimes called the Casimir operator,
$$
{\bf S}^2_{\rm tot}\, =\, (S^x_{\rm tot})^2 + (S^y_{\rm tot})^2 + (S^z_{\rm tot})^2\, .
$$
This is the generator of the center of the representation of ${\rm SU}(2)$;
therefore, total spin ${\bf S}^2_{\rm tot}$ commutes with the ``magnetization'' operator $S^z_{\rm tot}$.
Since the self-adjoint operators $S^z_{\rm tot}$ and ${\bf S}^2_{\rm tot}$ commute, there are simultaneous eigenspaces
$$
\Hil_{\rm tot}(s,m)\,
=\, \left \{ \psi \in \Hil_{\rm tot}\, :\,
{\bf S}^2_{\rm tot} \psi = s (s+1) \psi \ \text { and } \
S^z_{\rm tot} \psi = m \psi \right \}\, ,
$$
for $s \in \{0,1,\dots,\lfloor N/2 \rfloor\}$ and $m \in \{-s,-s+1,\dots,+s\}$.
It is here that we first use the fact that $N$ is even.
Writing $N=2n$, we have the direct sum decomposition
$$
\Hil_{\rm tot}\, =\, \bigoplus_{s=0}^{n} \bigoplus_{m=-s}^{s} \Hil_{\rm tot}(s,m)\, .
$$
The maximum possible value of $s$ is $n$, and similarly this is the maximum possible value of $m$.

Defining the spin raising and lowering operators
$$
S^{\pm}_{\rm tot}\, =\, S^x_{\rm tot} \pm i S^y_{\rm tot}\, ,
$$
these also commute with $H$.
Moreover, repeated applications of these raising and lowering operators may be used to map each space $\Hil_{\rm tot}(s,m)$ for a fixed value of $s$
onto all the other spaces with the same $s$ and different $m$-values.
Therefore, for a fixed value of $s$, the spectrum
$$
\operatorname{spec}(H \restriction \Hil_{\rm tot}(s,m))
$$
is the same for all $m$.
Considering an eigenvector $\psi \in \Hil_{\rm tot}(s,m)$, one often refers to the integer $n-m$ as the number
of ``magnons.''
This is the deviation of the magnetization eigenvalue from its maximum possible
value, which is $n$.
Equally important is the notion of ``spin deviation,'' which is $n-s$.
Let us denote
$$
E_0(C_{2n},k)\, 
=\, \min \operatorname{spec}( H \restriction \Hil_{\rm tot}(n-k,n-k))\quad \text { for } \quad k=0,\dots,n\, .
$$
The number $E_0(C_{2n},k)$ is the minimum energy on the length $N=2n$ cycle, among all eigenvectors 
with $k$ spin deviates.

We say {\em ferromagnetic ordering of energy levels} (FOEL) holds at level $k \in \{0,1,\dots,n\}$ for the cycle $C_{2n}$ if
$$
E_0(C_{2n},k)\, \leq\, E_0(C_{2n},\ell)\, ,
\quad \text { for all $\ell \in \{k,\dots,n\}$.}
$$
A famous result due to Lieb and Mattis establishes the opposite type of inequality
for antiferromagnets \cite{LiebMattis}.
Lieb and Mattis's result was known as ``ordering of energy levels,'' (OEL)
which is why we call the ferromagnetic version FOEL.

In \cite{NSS1}, we proved that the FOEL property holds at all levels whenever the underlying graph is an open chain
instead of a cycle: so the periodic edge $\{1,2n\}$ is absent.
This built on an earlier result of Koma and Nachtergaele, establishing FOEL at level $k=1$ for open chains.
Also, see \cite{NS,NNS} for results about FOEL on open chains for higher spin models.
The FOEL property for other groups has also been investigated by Hakobyan \cite{Hakobyan}.
The FOEL property, combined with Bethe's ansatz, is useful for understanding the low energy spectrum of the Heisenberg ferromagnet,
as well as the $q$-deformed $XXZ$ model in one-dimension, as we showed in \cite{NSS2}.

Moreover, FOEL at level $k=1$ is related to a conjecture due to Aldous \cite{Aldous} for the symmetric exclusion process,
whose Markov generator is unitarily equivalent to the spin-$1/2$ Heisenberg Hamiltonian.
Aldous's conjecture was recently proved by Caputo, Liggett and Richthammer \cite{CLR}.
More specifically, they proved a result which implies that FOEL holds at level $k=1$
for all finite graphs, not just open chains or cycles.
Prior to this, using a different technique, it was proved that FOEL holds at level $k=1$
for sufficiently large boxes \cite{ConomosStarr,Morris}.
In a forthcoming paper, it will be proved that FOEL holds at level $k$, for every $k$,
for sufficiently large boxes \cite{NSS3}.
Therefore, it is notable that we have found counterexamples to FOEL at level $k$
for $k=2,\dots,8$, which is the result of the present paper.
The method which proves FOEL at level $k>1$ only works for sufficiently large boxes,
but this is not just a deficiency of the proof technique because in fact FOEL
at level $k>1$ does not hold for all finite graphs.
This also gives some counterexamples to the more general property of ``Aldous ordering,''
introduced by Alon and Kozma in \cite{AK}.

\smallskip

{\em 
{\bf Main Results:} For each value of $n=2,3,\dots,8$, 
$$
E_0(C_{2n},n)\, \leq\, E_0(C_{2n},n-1)\, ,
$$
and there is strict inequality for $n>2$.}

\smallskip

For $n=2$ and $3$, we prove this result by algebraic diagonalization of the Hamiltonian operator.
That occupies Section 2.
In Section 3, we extend these results using exact numerical diagonalization, which establishes the result for $n=4,\dots,7$.
In Section 4, we discuss the relation to the Bethe ansatz, and also how to infer the result for $n=8$ from previous
results of Dhar and Shastry \cite{DharShastry}.

\section{Exact diagonalization of the graphical model}

For doing calculations, we find it convenient to use a graphical basis, related to the Temperley-Lieb algebra.
At first, we will consider the model on an open chain with $2n$ sites.
Later we consider how the periodic boundary conditions, and especially the edge $\{1,2n\}$
alter the set-up.

Starting from the all $\ket{\uparrow}$ spin state, one may make a $k$-spin deviate vector
as follows.
For each ordered set of $2k$ distinct sites $(a_1,b_1,\dots,a_k,b_k)$ define
$$
\Psi(a_1,b_1,\dots,a_k,b_k)\,
=\, \prod_{j=1}^{r} S^-_{a_j,b_j}\, \ket{\Uparrow}\, ,
$$
where $\ket{\Uparrow}$ is the all-up-spin state, and 
$$
S^-_{a,b}\, =\, S^-_{a} - S^-_{b}\, .
$$
We can write this instead as
$$
\Psi(a_1,b_1,\dots,a_k,b_k)\,
=\, \bigotimes_{j=1}^{k} \psi_{a_j,b_j} \otimes \bigotimes_{\ell \not\in \{a_1,b_1,\dots,a_k,b_k\}} \ket{\uparrow}_\ell\, ,
$$
where $\psi_{a,b} = \ket{\downarrow}_a \otimes \ket{\uparrow}_b - \ket{\uparrow}_a \otimes \ket{\downarrow}_b$.
There are some graphical rules, which are designed to obtain the right number of vectors this way to 
form a basis for $\Hil_{\rm tot}(n-k,n-k)$:
\begin{itemize}
\item For each $j$, $a_j<b_j$.
\item Each $j \in \{1,\dots,2n\}$ must appear in the list $(a_1,b_1,\dots,a_k,b_j)$ at most once.
If it does not occur at all it is called an ``unpaired site.''
\item Drawing arcs from $a_j$ to $b_j$ for $j=1,\dots,k$, no two arcs may cross.
\item No arc may span an unpaired site.
\end{itemize}
Graphically, we represent this as follows.
We draw a line of $2n$ dots.
Above the line, we draw an oriented arc from $a_j$ to $b_j$ for each $j$,
and we either draw the $\uparrow$ vectors above each unpaired spin site
or else we draw a 
$\begin{minipage}{5mm}
\begin{tikzpicture} 
\filldraw[fill=white] (2.5,1.5) circle (1.5mm);
\draw (2.5,1.5) node [] {\small $\boldsymbol{\uparrow}$}; 
\end{tikzpicture}
\end{minipage}$ 
above the line, indicating the $\uparrow$ spins, and draw strands
from each unpaired site to this:
\begin{align*}
\Psi(2,3;5,8;6,7)\quad
&=\quad
\begin{minipage}{7cm}
\begin{tikzpicture}
\draw[thick] (2,0) .. controls (2,0.5) and (3,0.5) .. (3,0);
\draw[thick] (6,0) .. controls (6,0.5) and (7,0.5) .. (7,0);
\draw[thick] (5,0) .. controls (5,1) and (8,1) .. (8,0);
\fill[white] (1,0) circle (1mm); \fill (1,0) circle (0.75mm);
\fill[white] (2,0) circle (1mm); \fill (2,0) circle (0.75mm);
\fill[white] (3,0) circle (1mm); \fill (3,0) circle (0.75mm);
\fill[white] (4,0) circle (1mm); \fill (4,0) circle (0.75mm);
\fill[white] (5,0) circle (1mm); \fill (5,0) circle (0.75mm);
\fill[white] (6,0) circle (1mm); \fill (6,0) circle (0.75mm);
\fill[white] (7,0) circle (1mm); \fill (7,0) circle (0.75mm);
\fill[white] (8,0) circle (1mm); \fill (8,0) circle (0.75mm);
\draw (1,0.1) node [above] {$\uparrow$};
\draw (4,0.1) node [above] {$\uparrow$};
\end{tikzpicture}
\end{minipage}\\[10pt]
&=\quad
\begin{minipage}{7cm}
\begin{tikzpicture}
\draw (1,0) .. controls (1,0.5) and (2,1) .. (2.5,1.5);
\draw (4,0) .. controls (4,0.5) and (3,1) .. (2.5,1.5);
\filldraw[fill=white] (2.5,1.5) circle (1.5mm);
\draw (2.5,1.5) node [] {\small $\boldsymbol{\uparrow}$};
\draw[thick] (2,0) .. controls (2,0.5) and (3,0.5) .. (3,0);
\draw[thick] (6,0) .. controls (6,0.5) and (7,0.5) .. (7,0);
\draw[thick] (5,0) .. controls (5,1) and (8,1) .. (8,0);
\fill[white] (1,0) circle (1mm); \fill (1,0) circle (0.75mm);
\fill[white] (2,0) circle (1mm); \fill (2,0) circle (0.75mm);
\fill[white] (3,0) circle (1mm); \fill (3,0) circle (0.75mm);
\fill[white] (4,0) circle (1mm); \fill (4,0) circle (0.75mm);
\fill[white] (5,0) circle (1mm); \fill (5,0) circle (0.75mm);
\fill[white] (6,0) circle (1mm); \fill (6,0) circle (0.75mm);
\fill[white] (7,0) circle (1mm); \fill (7,0) circle (0.75mm);
\fill[white] (8,0) circle (1mm); \fill (8,0) circle (0.75mm);
\end{tikzpicture}
\end{minipage}\quad .
\end{align*}
More generally, for an unpaired site, we could place an $\uparrow$
or a $\downarrow$ spin.
However, we get a basis vector in $\Hil_{\rm tot}(n-k,n-k)$ only if all unpaired
spins are $\uparrow$ spins.
Such a vector is called a ``highest weight'' vector because for total spin $s=n-k$
it has the highest possible value of the $S^z_{\rm tot}$ eigenvalue $m$, namely $n-k$.

Co-vectors, i.e., linear functionals, have a slightly different form.
We define
$$
\tilde{\psi}^\dagger(a,b)\, =\, {}_a\bra{\uparrow} \otimes {}_b\bra{\downarrow} - 
{}_b\bra{\uparrow} \otimes {}_a\bra{\downarrow}\, ,
$$
which has the opposite orientation so that,
$$
\tilde{\psi}^\dagger_{ab} \psi_{ab}\, =\, -2\, .
$$
This is convenient in the graphical representation for the following reason.
We denote $\tilde{\psi}^\dagger_{ab}$ as an arc between the sites $a$ and $b$.
We also denote the covector $\bra{\uparrow}$ or $\bra{\downarrow}$ at a site
by placing the spin configuration under a site.
Then we have the useful formulas:
$$
\begin{minipage}{2cm}
\begin{tikzpicture}
\draw[thick] (2,0) .. controls (2,0.5) and (3,0.5) .. (3,0);
\draw[thick] (1,0) .. controls (1,-0.5) and (2,-0.5) .. (2,0);
\fill[white] (1,0) circle (1mm); \fill (1,0) circle (0.75mm);
\fill[white] (2,0) circle (1mm); \fill (2,0) circle (0.75mm);
\fill[white] (3,0) circle (1mm); \fill (3,0) circle (0.75mm);
\draw (1,0.1) node [above] {$\sigma$};
\draw (3,-0.1) node [below] {$\sigma'$};
\end{tikzpicture}
\end{minipage} \qquad 
= \quad \delta_{\sigma,\sigma'}\quad
= \quad \begin{minipage}{2cm}
\begin{tikzpicture}
\fill[white] (1,0) circle (1mm); \fill (1,0) circle (0.75mm);
\draw (1,0.1) node [above] {$\sigma$};
\draw (1,-0.1) node [below] {$\sigma'$};
\end{tikzpicture}
\end{minipage}
$$
and
$$
\begin{minipage}{2cm}
\begin{tikzpicture}[xscale=-1]
\draw[thick] (2,0) .. controls (2,0.5) and (3,0.5) .. (3,0);
\draw[thick] (1,0) .. controls (1,-0.5) and (2,-0.5) .. (2,0);
\fill[white] (1,0) circle (1mm); \fill (1,0) circle (0.75mm);
\fill[white] (2,0) circle (1mm); \fill (2,0) circle (0.75mm);
\fill[white] (3,0) circle (1mm); \fill (3,0) circle (0.75mm);
\draw (1,0.1) node [above] {$\sigma$};
\draw (3,-0.1) node [below] {$\sigma'$};
\end{tikzpicture}
\end{minipage} \qquad 
= \quad \begin{minipage}{2cm}
\begin{tikzpicture}
\fill[white] (1,0) circle (1mm); \fill (1,0) circle (0.75mm);
\draw (1,0.1) node [above] {$\sigma$};
\draw (1,-0.1) node [below] {$\sigma'$};
\end{tikzpicture}
\end{minipage} \qquad .
$$
These formulas may be checked by expanding each singlet vector $\psi_{ab}$ and
covector $\tilde{\psi}^\dagger_{ab}$. 
The graphical interpretation is that one may contract the middle arcs to simplify
the diagram.
This is the first instance of a general procedure related to the graphical representation,
interpreting algebraic relations as topological simplifications.
This is also the idea behind the Temperley-Lieb algebra.

\subsection{Temperley-Lieb algebra}
We know that $\tilde{\psi}^\dagger_{ab} \psi_{ab}=-2$.
This is represented diagrammatically as saying that the ``loop'' evaluates to $-2$:
$$
\begin{minipage}{1cm}
\begin{tikzpicture}
\draw[thick] (1,0) .. controls (1,0.5) and (2,0.5) .. (2,0);
\draw[thick] (1,0) .. controls (1,-0.5) and (2,-0.5) .. (2,0);
\fill[white] (1,0) circle (1mm); \fill (1,0) circle (0.75mm);
\fill[white] (2,0) circle (1mm); \fill (2,0) circle (0.75mm);
\end{tikzpicture}
\end{minipage} \qquad 
=\ -2\, .
$$
This is the scalar inner-product.
But we may also take the outer product to obtain an operator or endomorphism.
The Temperley-Lieb generators are defined as $U_1,\dots,U_{2n-1}$, where
$$
U_k\, =\, \unity_{\C^2}^{k-1} \otimes (\psi_{k,k+1} \tilde{\psi}^\dagger_{k,k+1}) \otimes 
\unity_{\C^2}^{2n-1-k}\, .
$$
One may check that
$$
U_k\, =\, -2 h_k\, ,
$$
for each $k\in \{1,\dots,2n-1\}$.
In order to represent these graphically, we represent a factor of $\unity_{\C^2}$ as a straight
vertical line segment, with the interpretation that contracting the line segment does implement
the Kronecker delta, as it should:
$$
U_{k}\, =\quad 
\begin{minipage}{8cm}
\begin{tikzpicture}[xscale=0.5,yscale=0.5]
\draw[very thick] (0,0) -- (0,-3);
\fill[white] (0,0) circle (5pt); \fill (0,0) circle (3pt);
\fill[white] (0,-3) circle (5pt); \fill (0,-3) circle (3pt) node[below] {$1$};
\draw (2,-1.5) node {\large \dots};
\draw[very thick] (4,0) -- (4,-3);
\fill[white] (4,0) circle (5pt); \fill (4,0) circle (3pt);
\fill[white] (4,-3) circle (5pt); \fill (4,-3) circle (3pt) node[below] {$k-1$};
\draw[very thick] (6,0) arc (180:360:1cm);
\fill[white] (6,0) circle (5pt); \fill (6,0) circle (3pt);
\fill[white] (8,0) circle (5pt); \fill (8,0) circle (3pt);
\draw[very thick] (6,-3) arc (180:0:1cm);
\fill[white] (6,-3) circle (5pt); \fill (6,-3) circle (3pt) node[below] {$k$};
\fill[white] (8,-3) circle (5pt); \fill (8,-3) circle (3pt) node[below] {$k+1$};
\draw[very thick] (10,0) -- (10,-3);
\fill[white] (10,0) circle (5pt); \fill (10,0) circle (3pt);
\fill[white] (10,-3) circle (5pt); \fill (10,-3) circle (3pt) node[right=0.5cm, below] {$k+2$};
\draw (12,-1.5) node {\large \dots};
\draw[very thick] (14,0) -- (14,-3);
\fill[white] (14,0) circle (5pt); \fill (14,0) circle (3pt);
\fill[white] (14,-3) circle (5pt); \fill (14,-3) circle (3pt) node[below] {$2n$};
\end{tikzpicture}
\end{minipage}
$$
for $k=1,\dots,2n-1$. 
The Temperley-Lieb algebra is the algebra obtained from these generators, using the relations
\begin{gather*}
U_{k}^2\, =\, -2 U_{k} \\[10pt]
U_{k} U_{k+1} U_{k}\, =\, U_{k}\\[10pt]
U_{k} U_{k-1} U_{k}\, =\, U_{k}\\[10pt]
|k-\ell| > 1 \ \Rightarrow\ U_{k} U_{\ell} = U_{\ell} U_{k}
\end{gather*}
With the representation given above, where $U_k = -2h_k$, these relations are all satisfied.
The graphical interpretations of these relations are as follows, where we leave out unnecessary
tensor factors of the identity to remove extraneous vertical lines:
$$
\begin{minipage}{1cm}
\begin{tikzpicture}[xscale=0.35,yscale=0.35]
\draw[very thick] (0,0) arc(180:360:1cm);
\draw[very thick] (0,-2.5) arc(180:0:1cm);
\draw[very thick] (0,-2.5) arc(180:360:1cm);
\draw[very thick] (0,-5) arc(180:0:1cm);
\end{tikzpicture}
\end{minipage}\
=\quad -2\
\begin{minipage}{1cm}
\begin{tikzpicture}[xscale=0.35,yscale=0.35]
\draw[very thick] (0,0) arc(180:360:1cm);
\draw[very thick] (0,-2.5) arc(180:0:1cm);
\end{tikzpicture}
\end{minipage}\ ,\qquad
\begin{minipage}{2cm}
\begin{tikzpicture}[xscale=0.35,yscale=0.35]
\draw[very thick] (0,0) arc(180:360:1cm);
\draw[very thick] (0,-2.5) arc(180:0:1cm);
\draw[very thick] (4,0) -- (4,-2.5);
\draw[very thick] (0,-2.5) -- (0,-5);
\draw[very thick] (2,-2.5) arc(180:360:1cm);
\draw[very thick] (2,-5) arc(180:0:1cm);
\draw[very thick] (0,-5) arc(180:360:1cm);
\draw[very thick] (0,-7.5) arc(180:0:1cm);
\draw[very thick] (4,-5) -- (4,-7.5);
\end{tikzpicture}
\end{minipage}\
=\quad
\begin{minipage}{2cm}
\begin{tikzpicture}[xscale=0.35,yscale=0.35]
\draw[very thick] (0,0) arc(180:360:1cm);
\draw[very thick] (0,-2.5) arc(180:0:1cm);
\draw[very thick] (4,0) -- (4,-2.5);
\end{tikzpicture}
\end{minipage}\quad ,
$$
and
$$
\begin{minipage}{1.75cm}
\begin{tikzpicture}[xscale=-0.25,yscale=0.25]
\draw[very thick] (0,0) arc(180:360:1cm);
\draw[very thick] (0,-2.5) arc(180:0:1cm);
\draw[very thick] (4,0) -- (4,-2.5);
\draw[very thick] (0,-2.5) -- (0,-5);
\draw[very thick] (2,-2.5) arc(180:360:1cm);
\draw[very thick] (2,-5) arc(180:0:1cm);
\draw[very thick] (0,-5) arc(180:360:1cm);
\draw[very thick] (0,-7.5) arc(180:0:1cm);
\draw[very thick] (4,-5) -- (4,-7.5);
\end{tikzpicture}
\end{minipage}\
=\quad
\begin{minipage}{1.75cm}
\begin{tikzpicture}[xscale=-0.25,yscale=0.25]
\draw[very thick] (0,0) arc(180:360:1cm);
\draw[very thick] (0,-2.5) arc(180:0:1cm);
\draw[very thick] (4,0) -- (4,-2.5);
\end{tikzpicture}
\end{minipage}\quad ,\qquad
\begin{minipage}{2.5cm}
\begin{tikzpicture}[xscale=0.25,yscale=0.25]
\draw[very thick] (0,0) arc(180:360:1cm);
\draw[very thick] (0,-2.5) arc(180:0:1cm);
\draw[very thick] (4,0) -- (4,-2.5);
\draw[very thick] (6,0) -- (6,-2.5);
\draw[very thick] (0,-2.5) -- (0,-5);
\draw[very thick] (2,-2.5) -- (2,-5);
\draw[very thick] (4,-2.5) arc(180:360:1cm);
\draw[very thick] (4,-5) arc(180:0:1cm);
\end{tikzpicture}
\end{minipage}\
=\quad
\begin{minipage}{2.25cm}
\begin{tikzpicture}[xscale=-0.25,yscale=0.25]
\draw[very thick] (0,0) arc(180:360:1cm);
\draw[very thick] (0,-2.5) arc(180:0:1cm);
\draw[very thick] (4,0) -- (4,-2.5);
\draw[very thick] (6,0) -- (6,-2.5);
\draw[very thick] (0,-2.5) -- (0,-5);
\draw[very thick] (2,-2.5) -- (2,-5);
\draw[very thick] (4,-2.5) arc(180:360:1cm);
\draw[very thick] (4,-5) arc(180:0:1cm);
\end{tikzpicture}
\end{minipage} .
$$
These may also be interpreted as topological simplifications.
The model for many of these algebraic formalisms of topological simplifications
is given by Reidemeister's ``moves'' for knot and braid diagrams.
Indeed, the Temperley-Lieb algebra played an important role in the discovery by Vaughn Jones
of his knot polynomial.
We recommend the article of Frenkel and Khovanov for the description
in the setting of the quantum group ${\rm SU}_q(2)$ \cite{FK}.
The graphical representation has been important in proving FOEL for open
chains, as in \cite{NSS1} and more recently \cite{NNS}.

Since $U_k = -2h_k$, for $k=1,\dots,2n-1$, this gives a convenient representation for the Hamiltonian.
One may consider the formal vector space spanned by the diagrams for vectors, and use an operator $A$
representing the sum of the Temperley-Lieb generators.
Then, after finding the eigenvalues and eigenvectors in this basis, one may transform back into the standard
(Ising) tensor basis of $\ket{\uparrow}$ and $\ket{\downarrow}$ vectors by expanding each $\psi_{ab}$ singlet
vector.
As some examples of the diagrammatic rules for the graphical representations of $U_k$, we consider the following
\begin{gather*}
U_{2} \left(\psi_{12}\otimes \bigotimes_{j=3,4} \ket{\uparrow}_j\right)\, 
=\quad
\begin{minipage}{3cm}
\begin{tikzpicture}
\draw (3,0) .. controls (3,0.5) and (3.25,0.5) .. (3.5,1);
\draw (4,0) .. controls (4,0.5) and (3.75,0.5) .. (3.5,1);
\filldraw[fill=white] (3.5,1) circle (1.5mm);
\draw (3.5,1) node [] {\small $\boldsymbol{\uparrow}$};
\draw[thick] (1,0) .. controls (1,0.5) and (2,0.5) .. (2,0);
\draw[thick] (1,0) -- (1,-1);
\draw[thick] (4,0) -- (4,-1);
\draw[thick] (2,0) .. controls (2,-0.5) and (3,-0.5) .. (3,0);
\draw[thick] (2,-1) .. controls (2,-0.5) and (3,-0.5) .. (3,-1);
\fill[white] (1,0) circle (1mm); \fill (1,0) circle (0.75mm);
\fill[white] (2,0) circle (1mm); \fill (2,0) circle (0.75mm);
\fill[white] (3,0) circle (1mm); \fill (3,0) circle (0.75mm);
\fill[white] (4,0) circle (1mm); \fill (4,0) circle (0.75mm);
\fill[white] (1,-1) circle (1mm); \fill (1,-1) circle (0.75mm);
\fill[white] (2,-1) circle (1mm); \fill (2,-1) circle (0.75mm);
\fill[white] (3,-1) circle (1mm); \fill (3,-1) circle (0.75mm);
\fill[white] (4,-1) circle (1mm); \fill (4,-1) circle (0.75mm);
\end{tikzpicture}
\end{minipage} \qquad
%=\quad
%\begin{minipage}{3cm}
%\begin{tikzpicture}
%\draw (1,0) .. controls (1,0.5) and (2,0.75) .. (2.5,1.25);
%\draw (4,0) .. controls (4,0.5) and (3,0.75) .. (2.5,1.25);
%\filldraw[fill=white] (2.5,1.25) circle (1.5mm);
%\draw (2.5,1.25) node [] {\small $\boldsymbol{\uparrow}$};
%\draw[thick] (2,0) .. controls (2,0.5) and (3,0.5) .. (3,0);
%\fill[white] (1,0) circle (1mm); \fill (1,0) circle (0.75mm);
%\fill[white] (2,0) circle (1mm); \fill (2,0) circle (0.75mm);
%\fill[white] (3,0) circle (1mm); \fill (3,0) circle (0.75mm);
%\fill[white] (4,0) circle (1mm); \fill (4,0) circle (0.75mm);
%\end{tikzpicture}
%\end{minipage}\qquad
=\ \psi_{23}\otimes \bigotimes_{j=1,4} \ket{\uparrow}_j\, ,\\[10pt]
U_{2} \left(\psi_{23}\otimes \bigotimes_{j=1,4} \ket{\uparrow}_j\right)\, 
=\quad
\begin{minipage}{3cm}
\begin{tikzpicture}
\draw (1,0) .. controls (1,0.5) and (2,0.75) .. (2.5,1.25);
\draw (4,0) .. controls (4,0.5) and (3,0.75) .. (2.5,1.25);
\filldraw[fill=white] (2.5,1.25) circle (1.5mm);
\draw (2.5,1.25) node [] {\small $\boldsymbol{\uparrow}$};
\draw[thick] (2,0) .. controls (2,0.5) and (3,0.5) .. (3,0);
\draw[thick] (1,0) -- (1,-1);
\draw[thick] (4,0) -- (4,-1);
\draw[thick] (2,0) .. controls (2,-0.5) and (3,-0.5) .. (3,0);
\draw[thick] (2,-1) .. controls (2,-0.5) and (3,-0.5) .. (3,-1);
\fill[white] (1,0) circle (1mm); \fill (1,0) circle (0.75mm);
\fill[white] (2,0) circle (1mm); \fill (2,0) circle (0.75mm);
\fill[white] (3,0) circle (1mm); \fill (3,0) circle (0.75mm);
\fill[white] (4,0) circle (1mm); \fill (4,0) circle (0.75mm);
\fill[white] (1,-1) circle (1mm); \fill (1,-1) circle (0.75mm);
\fill[white] (2,-1) circle (1mm); \fill (2,-1) circle (0.75mm);
\fill[white] (3,-1) circle (1mm); \fill (3,-1) circle (0.75mm);
\fill[white] (4,-1) circle (1mm); \fill (4,-1) circle (0.75mm);
\end{tikzpicture}
\end{minipage} \qquad
%=\quad -2\quad 
%\begin{minipage}{3cm}
%\begin{tikzpicture}
%\draw (1,0) .. controls (1,0.5) and (2,0.75) .. (2.5,1.25);
%\draw (4,0) .. controls (4,0.5) and (3,0.75) .. (2.5,1.25);
%\filldraw[fill=white] (2.5,1.25) circle (1.5mm);
%\draw (2.5,1.25) node [] {\small $\boldsymbol{\uparrow}$};
%\draw[thick] (2,0) .. controls (2,0.5) and (3,0.5) .. (3,0);
%\fill[white] (1,0) circle (1mm); \fill (1,0) circle (0.75mm);
%\fill[white] (2,0) circle (1mm); \fill (2,0) circle (0.75mm);
%\fill[white] (3,0) circle (1mm); \fill (3,0) circle (0.75mm);
%\fill[white] (4,0) circle (1mm); \fill (4,0) circle (0.75mm);
%\end{tikzpicture}
%\end{minipage}\qquad
=\ -2 \psi_{23}\otimes \bigotimes_{j=1,4} \ket{\uparrow}_j\, ,\\[10pt]
U_{2} \left(\psi_{12}\otimes \psi_{34}\right)\, 
=\quad
\begin{minipage}{3cm}
\begin{tikzpicture}
\draw[thick] (1,0) .. controls (1,0.5) and (2,0.5) .. (2,0);
\draw[thick] (3,0) .. controls (3,0.5) and (4,0.5) .. (4,0);
\draw[thick] (1,0) -- (1,-1);
\draw[thick] (4,0) -- (4,-1);
\draw[thick] (2,0) .. controls (2,-0.5) and (3,-0.5) .. (3,0);
\draw[thick] (2,-1) .. controls (2,-0.5) and (3,-0.5) .. (3,-1);
\fill[white] (1,0) circle (1mm); \fill (1,0) circle (0.75mm);
\fill[white] (2,0) circle (1mm); \fill (2,0) circle (0.75mm);
\fill[white] (3,0) circle (1mm); \fill (3,0) circle (0.75mm);
\fill[white] (4,0) circle (1mm); \fill (4,0) circle (0.75mm);
\fill[white] (1,-1) circle (1mm); \fill (1,-1) circle (0.75mm);
\fill[white] (2,-1) circle (1mm); \fill (2,-1) circle (0.75mm);
\fill[white] (3,-1) circle (1mm); \fill (3,-1) circle (0.75mm);
\fill[white] (4,-1) circle (1mm); \fill (4,-1) circle (0.75mm);
\end{tikzpicture}
\end{minipage} \qquad
%=\quad
%\begin{minipage}{3cm}
%\begin{tikzpicture}
%\draw[thick] (1,0) .. controls (1.5,1) and (3.5,1) .. (4,0);
%\draw[thick] (2,0) .. controls (2,0.5) and (3,0.5) .. (3,0);
%\fill[white] (1,0) circle (1mm); \fill (1,0) circle (0.75mm);
%\fill[white] (2,0) circle (1mm); \fill (2,0) circle (0.75mm);
%\fill[white] (3,0) circle (1mm); \fill (3,0) circle (0.75mm);
%\fill[white] (4,0) circle (1mm); \fill (4,0) circle (0.75mm);
%\end{tikzpicture}
%\end{minipage}\qquad
=\ \psi_{14}\otimes \psi_{23}\, ,
\end{gather*}
Also, since the singlet is antisymmetric, while $\ket{\uparrow} \otimes \ket{\uparrow}$ is clearly symmetric,
this implies
$$
U_1 (\ket{\uparrow}_1 \otimes \ket{\uparrow}_2)\
=\qquad 
\begin{minipage}{1cm}
\begin{tikzpicture}
\draw[thick] (1,0) .. controls (1,0.5) and (1.25,0.5) .. (1.5,1);
\draw[thick] (2,0) .. controls (2,0.5) and (1.75,0.5) .. (1.5,1);
\filldraw[fill=white] (1.5,1) circle (1.5mm);
\draw (1.5,1) node [] {\small $\boldsymbol{\uparrow}$};
\draw[thick] (1,0) .. controls (1,-0.5) and (2,-0.5) .. (2,0);
\fill[white] (1,0) circle (1mm); \fill (1,0) circle (0.75mm);
\fill[white] (2,0) circle (1mm); \fill (2,0) circle (0.75mm);
\end{tikzpicture}
\end{minipage}\qquad
=\ 0\, .
$$
A more general identity of this sort involves the Jones-Wenzl projector, which is a  symmetrizer
for two or more spins.
We will not need the more general identity.

\subsection{Graphical representation on the ring}

We want to consider the graphical representation on the ring.
However, momentarily let us continue to 
consider the spin-$1/2$ Heisenberg model on the chain, so that the edge $\{1,2n\}$ is still absent,
as is the interaction $h_{2n}$.

For the action on the $k$ spin deviate subspace, we may define another
vector space $V$ which is the set of formal linear combinations of the arc-diagrams on the chain.
Then we may define an operator $A$ acting on $V$, obeying the appropriate
rules one infers from the Temperley-Lieb algebra.
This will be the sum of all Temperly-Lieb generators:
$$
A\, =\, U_{1,2} + \dots + U_{2n-1,2n}\, .
$$
(We now write $U_k$ as $U_{k,k+1}$ to emphasize both vertices involved in the arcs.)
Finally we may define a linear transformation between vector spaces,
$$
L : V \to \Hil_{\rm tot}(n-k,n-k)\, ,
$$ 
which implements
the original definition of the arc diagrams in terms of linear combinations of Ising spin configuration
basis vectors.
Because the Temperley-Lieb algebra is related to the Heisenberg model by the identities $U_{k,k+1} = -2h_k$,
for $k=1,\dots,2n-1$, 
we then have the intertwining relation
$$
LA\, =\, -2HL\, .
$$
Therefore, if there is some eigenvector $\phi$ of $A$, such that $A \phi = \lambda \phi$,
then the vector $\psi = L \phi$ must satisfy $H \psi = -(1/2)\lambda \psi$.
This shows that $-(1/2)\lambda$ is an eigenvector of $H$, as long as
$\psi = L \phi$ is not zero.
For the case we are considering, $L$ is actually an isomorphism, as can be seen by counting
the number of basis vectors.
See, for example, \cite{NSS1} or \cite{NS} for a proof of the combinatorial identity
showing that the dimensions of $V$ and $\Hil_{\rm tot}(n-k,n-k)$ match.

So it never happens that $L\phi = 0$ if $\phi$ is nonzero.
Therefore, the entire spectrum of $(H \restriction \Hil_{\rm tot}(n-k,n-k))$ can
be obtained merely by multiplying all points of $\operatorname{spec}(A)$ by $-1/2$.

In order to consider the cycle we now need to include $h_{2n}$.
This is not a simple Temperley-Lieb algebra generator.
We note for example, that the singlet vector satisfies
$$
\psi_{2n,1}\, =\, -(\psi_{12} + \psi_{23} + \dots + \psi_{2n-1,2n})\, ,
$$
with a similar formula for the covector.
But in the graphical model, we may merely add another generator
which acts in the correct way, according to the rules of the Templerley-Lieb algebra.

In other words, instead of decomposing an arc such as that representing $\psi_{2n,1}$
as a linear combination of the other more basic arc diagrams, we merely append that arc as part of the vector space $V$.
Thus we enlarge the vector space $V$, and therefore the intertwining operator 
$L : V \to \Hil_{\rm tot}(n-k,n-k)$ will no longer be an isomorphism.
It will map onto $\Hil_{\rm tot}(n-k,n-k)$, but $L$ will
have a kernel.
Taking this into account, and defining $A = U_{12} + \dots U_{2n-1,2n} + U_{2n,1}$, we have
$$
\operatorname{spec}(H \restriction \Hil_{\rm tot}(n-k,n-k))\,
=\, \{-(1/2) \lambda\, :\, \exists \phi \in V\, ,\ A\phi = \lambda \phi \quad \text{ and } \quad \phi \not\in \ker(L)\}\, .
$$

When working on the cycle, we must specify the orientation of each arc,
because it is not uniquely determined as left-to-right as for the chain.
Reversing the orientation of a single arc corresponds to multiplying a vector in $\Hil_{\rm}(n-k,n-k)$ by
$-1$, 
after the action of $L$.
Note, however, that reversing two arcs does not change the sign.
For this reason, in a Temperley-Lieb generator $U_{k,k+1}$ reversing the orientation of both
arcs, simultaneously, does not change its action.

In principle we could allow $V$ to be the formal linear combination of all oriented arc diagrams.
However, we will typically choose a subset.
For each arc configuration, we will want to consider all translates of it, under the rotation group
which is a symmetry of the operator $A$ (and $H$).
This simplifies the calculations.
However, we often times do not do the same for the reflections, since the full dihedral symmetry $D_{2n}$
is not much more useful than the cyclic symmetry $C_{2n}$.

\subsection{The square $C_4$}

We begin by considering $C_4$ and $k=1$.

\subsubsection{$C_4$, 1 spin deviate}
\label{subsec:C4sd1}
We consider $V$ to be the formal span of the following diagrams.
$$
\phi_{12}\, =\quad
\begin{minipage}{1.5cm}
\begin{tikzpicture}
\draw[very thick] (0,0) circle (7.5mm);
\fill (45:7.5mm) circle (0.75mm);
\fill (315:7.5mm) circle (0.75mm);
\fill (225:7.5mm) circle (0.75mm);
\fill (135:7.5mm) circle (0.75mm);
\draw[thick] (45:7.5mm) .. controls (45:3mm) and (315:3mm) .. (315:7.5mm);
\fill (0:3mm) +(0,0) -- +(-0.1,0.1) -- +(0.1,0.1);
\draw (225:7.5mm) -- (-0.1,0);
\draw (135:7.5mm) -- (-0.1,0);
\filldraw[fill=white] (-0.1,0) circle (1.5mm);
\draw (-0.1,0) node[] {\small $\boldsymbol{\uparrow}$};
\end{tikzpicture}
\end{minipage}\quad ,\qquad
\phi_{23}\, =\quad 
\begin{minipage}{1.5cm}
\begin{tikzpicture}[rotate=-90]
\draw[very thick] (0,0) circle (7.5mm);
\fill (45:7.5mm) circle (0.75mm);
\fill (315:7.5mm) circle (0.75mm);
\fill (225:7.5mm) circle (0.75mm);
\fill (135:7.5mm) circle (0.75mm);
\draw[thick] (45:7.5mm) .. controls (45:3mm) and (315:3mm) .. (315:7.5mm);
\fill (0:3mm) +(0,0) -- +(-0.1,0.1) -- +(0.1,0.1);
\draw (225:7.5mm) -- (-0.1,0);
\draw (135:7.5mm) -- (-0.1,0);
\filldraw[fill=white] (-0.1,0) circle (1.5mm);
\draw (-0.1,0) node[] {\small $\boldsymbol{\uparrow}$};
\end{tikzpicture}
\end{minipage}\quad , \qquad
\phi_{34}\, =\quad
\begin{minipage}{1.5cm}
\begin{tikzpicture}[rotate=180]
\draw[very thick] (0,0) circle (7.5mm);
\fill (45:7.5mm) circle (0.75mm);
\fill (315:7.5mm) circle (0.75mm);
\fill (225:7.5mm) circle (0.75mm);
\fill (135:7.5mm) circle (0.75mm);
\draw[thick] (45:7.5mm) .. controls (45:3mm) and (315:3mm) .. (315:7.5mm);
\fill (0:3mm) +(0,0) -- +(-0.1,0.1) -- +(0.1,0.1);
\draw (225:7.5mm) -- (-0.1,0);
\draw (135:7.5mm) -- (-0.1,0);
\filldraw[fill=white] (-0.1,0) circle (1.5mm);
\draw (-0.1,0) node[] {\small $\boldsymbol{\uparrow}$};
\end{tikzpicture}
\end{minipage}\quad ,
$$
and
$$
\phi_{41}\, =\quad 
\begin{minipage}{1.5cm}
\begin{tikzpicture}[rotate=90]
\draw[very thick] (0,0) circle (7.5mm);
\fill (45:7.5mm) circle (0.75mm);
\fill (315:7.5mm) circle (0.75mm);
\fill (225:7.5mm) circle (0.75mm);
\fill (135:7.5mm) circle (0.75mm);
\draw[thick] (45:7.5mm) .. controls (45:3mm) and (315:3mm) .. (315:7.5mm);
\fill (0:3mm) +(0,0) -- +(-0.1,0.1) -- +(0.1,0.1);
\draw (225:7.5mm) -- (-0.1,0);
\draw (135:7.5mm) -- (-0.1,0);
\filldraw[fill=white] (-0.1,0) circle (1.5mm);
\draw (-0.1,0) node[] {\small $\boldsymbol{\uparrow}$};
\end{tikzpicture}
\end{minipage}\quad .
$$
We enumerate the vertices from $1$ to $4$ starting at the top right, and proceeding clockwise.
We let $T$ be the rotation one unit to the right, so that
$$
\begin{bmatrix} \phi_{12} \\ \phi_{23} \\ \phi_{34} \\ \phi_{41} \end{bmatrix}\,
=\, \begin{bmatrix} 1 \\ T \\ T^2 \\ T^3 \end{bmatrix} \phi_{12}\, .
$$
Then, letting $U_{12}$, $U_{23}$, $U_{34}$ and $U_{41}$ be the four generators in the extended 
Temperley-Lieb algebra on the ring,
we have
$$
\begin{bmatrix}
U_{12}\\
U_{23}\\
U_{34}\\
U_{41}
\end{bmatrix}
 \phi_{12}\,
=\, \begin{bmatrix} -2 \\ T \\ 0 \\ T^3 \end{bmatrix} \phi_{12}\, .
$$
Therefore, for 
$$
A\, =\, U_{12} + U_{23} + U_{34} + U_{41}\, ,
$$
we have $A \phi_{12} = (-2+T+T^3) \phi_{12}$.
We define translation eigenvectors next.
For $z \in \{1,i,-1,-i\}$, we define
$$
\hat{\phi}(z)\, =\, \sum_{k=0}^{3} z^k T^{k} \phi_{12}\, .
$$
These are also eigenvectors of $A$:
\begin{equation}
\label{eq:Aspec1}
A \hat{\phi}(z)\, =\, (-2+z+z^3) \hat{\phi}(z) \qquad \Rightarrow \qquad
\left(A - 
\begin{bmatrix} 
0 \\
& -2 \\
& & -2 \\
& & & -4
\end{bmatrix}\right)
\begin{bmatrix}
\hat{\phi}(1)\\
\hat{\phi}(i)\\
\hat{\phi}(-i)\\
\hat{\phi}(-1)
\end{bmatrix}\,
=\, \begin{bmatrix}
0 \\ 0 \\ 0 \\ 0 
\end{bmatrix}\, .
\end{equation}

Next we must calculate the transformation to the $1$-spin deviate subspace for $C_4$,
$$
L\, :\, V \to \Hil_{\rm tot}(1,1)\, .
$$
We denote one-magnon Ising basis vectors in $\Hil_{\rm tot}$ for $C_4$ as
$$
\Psi_{1}\, =\,  \begin{minipage}{1.5cm}
\begin{tikzpicture}
\draw [] (0,0) circle (7.5mm);
\fill (45:7.5mm) circle (0.75mm);
\fill (135:7.5mm) circle (0.75mm);
\fill (225:7.5mm) circle (0.75mm);
\fill (315:7.5mm) circle (0.75mm);
\draw (45:1cm) node[] {$\downarrow$};
\draw (135:1cm) node[] {$\uparrow$};
\draw (225:1cm) node[] {$\uparrow$};
\draw (315:1cm) node[] {$\uparrow$};
\end{tikzpicture}
\end{minipage}\qquad ,\qquad
\Psi_{2}\, =\,  \begin{minipage}{1.5cm}
\begin{tikzpicture}[rotate=-90]
\draw [] (0,0) circle (7.5mm);
\fill (45:7.5mm) circle (0.75mm);
\fill (135:7.5mm) circle (0.75mm);
\fill (225:7.5mm) circle (0.75mm);
\fill (315:7.5mm) circle (0.75mm);
\draw (45:1cm) node[] {$\downarrow$};
\draw (135:1cm) node[] {$\uparrow$};
\draw (225:1cm) node[] {$\uparrow$};
\draw (315:1cm) node[] {$\uparrow$};
\end{tikzpicture}
\end{minipage}\qquad ,\qquad
\Psi_{3}\, =\,  \begin{minipage}{1.5cm}
\begin{tikzpicture}[rotate=180]
\draw [] (0,0) circle (7.5mm);
\fill (45:7.5mm) circle (0.75mm);
\fill (135:7.5mm) circle (0.75mm);
\fill (225:7.5mm) circle (0.75mm);
\fill (315:7.5mm) circle (0.75mm);
\draw (45:1cm) node[] {$\downarrow$};
\draw (135:1cm) node[] {$\uparrow$};
\draw (225:1cm) node[] {$\uparrow$};
\draw (315:1cm) node[] {$\uparrow$};
\end{tikzpicture}
\end{minipage}\qquad ,
$$
and
$$
\Psi_{4}\, =\,  \begin{minipage}{1.5cm}
\begin{tikzpicture}[rotate=90]
\draw [] (0,0) circle (7.5mm);
\fill (45:7.5mm) circle (0.75mm);
\fill (135:7.5mm) circle (0.75mm);
\fill (225:7.5mm) circle (0.75mm);
\fill (315:7.5mm) circle (0.75mm);
\draw (45:1cm) node[] {$\downarrow$};
\draw (135:1cm) node[] {$\uparrow$};
\draw (225:1cm) node[] {$\uparrow$};
\draw (315:1cm) node[] {$\uparrow$};
\end{tikzpicture}
\end{minipage}\qquad .
$$
We let $T$ also denote the implementation of the same rotation
operation on $\Hil_{\rm tot}$, through its Ising basis.
Then 
$$
\begin{bmatrix} \Psi_{1} \\ \Psi_{2} \\ \Psi_{3} \\ \Psi_{4} \end{bmatrix}\,
=\, \begin{bmatrix} 1 \\ T \\ T^2 \\ T^3 \end{bmatrix} \Psi_{1}\, .
$$
Also,
\begin{equation}
\label{eq:Lfirst}
L \phi_{12}\, =\, \Psi_1 - \Psi_2\, =\, (1-T) \Psi_1\, .
\end{equation}
We define translation eigenvectors in $\Hil_{\rm tot}$.
For $z \in \{1,i,-1,-i\}$, we define
$$
\hat{\psi}(z)\, =\, \sum_{k=0}^{3} z^k T^k \Psi_1\, .
$$
These are linearly independent vectors.
Moreover, from (\ref{eq:Lfirst}), we have
$$
L \hat{\phi}(z)\, =\, (1-z^3) \hat{\psi}(z)\, .
$$
Therefore the kernel of $L$ is one dimensional, spanned just by $\hat{\phi}(1)$.
So, using (\ref{eq:Aspec1}), we see that the spectrum of $H$ on the 1 spin deviate subspace for $C_4$ is given by
$$
\operatorname{spec}(2H \restriction \Hil_{\rm tot}(1,1))\, =\, \{2,4\}\, .
$$
In particular,
\begin{equation}
\label{eq:E0first}
E_0(C_4,1)\, =\, 1\, .
\end{equation}

\subsubsection{$C_4$, 2 spin deviate}

We consider the formal vector space $V$, spanned by two vectors
$$
\phi_{12,34}\, =\quad
\begin{minipage}{1.5cm}
\begin{tikzpicture}
\draw[very thick] (0,0) circle (7.5mm);
\fill (45:7.5mm) circle (0.75mm);
\fill (315:7.5mm) circle (0.75mm);
\fill (225:7.5mm) circle (0.75mm);
\fill (135:7.5mm) circle (0.75mm);
\draw[thick] (45:7.5mm) .. controls (45:3mm) and (315:3mm) .. (315:7.5mm);
\fill (0:3mm) +(0,0) -- +(-0.1,0.1) -- +(0.1,0.1);
\draw[thick] (225:7.5mm) .. controls (225:3mm) and (135:3mm) .. (135:7.5mm);
\fill (180:3mm) +(0,0) -- +(-0.1,-0.1) -- +(0.1,-0.1);
\end{tikzpicture}
\end{minipage}\quad ,\qquad
\phi_{14,23}\, =\quad 
\begin{minipage}{1.5cm}
\begin{tikzpicture}[rotate=-90]
\draw[very thick] (0,0) circle (7.5mm);
\fill (45:7.5mm) circle (0.75mm);
\fill (315:7.5mm) circle (0.75mm);
\fill (225:7.5mm) circle (0.75mm);
\fill (135:7.5mm) circle (0.75mm);
\draw[thick] (45:7.5mm) .. controls (45:3mm) and (315:3mm) .. (315:7.5mm);
\fill (0:3mm) +(0,0) -- +(-0.1,0.1) -- +(0.1,0.1);
\draw[thick] (225:7.5mm) .. controls (225:3mm) and (135:3mm) .. (135:7.5mm);
\fill (180:3mm) +(0,0) -- +(-0.1,-0.1) -- +(0.1,-0.1);
\end{tikzpicture}
\end{minipage}\quad .
$$
Then
$$
\begin{bmatrix}
\phi_{12,34}\\ 
\phi_{14,23}
\end{bmatrix}\,
=\,
\begin{bmatrix}
1 \\ 
T
\end{bmatrix}
\phi_{12,34}\, .
$$
Also,
$$
\begin{bmatrix} U_{12} \\ U_{23} \\ U_{34} \\ U_{41} \end{bmatrix} \phi_{12,34}\,
=\, \begin{bmatrix} -2 \\ T \\ -2 \\ T \end{bmatrix} \phi_{12,34}\, .
$$
We define the translation eigenvectors, for $z \in \{1,-1\}$,
$$
\hat{\phi}(z)\, =\, \sum_{k=0}^{1} z^k T^{k} \phi_{12,34}\, .
$$
Then for $A = U_{12} + U_{23} +U_{34} + U_{41}$, we have
\begin{equation}
\label{eq:Aspec2}
A \hat{\phi}(z)\, =\, (-4+2z) \hat{\phi}(z)\qquad \Rightarrow \qquad
\left(A - \begin{bmatrix} -2 \\ & -6 \end{bmatrix}\right)
\begin{bmatrix} \hat{\phi}(1) \\ \hat{\phi}(-1) \end{bmatrix}\,
=\, \begin{bmatrix} 0 \\ 0 \end{bmatrix}\, .
\end{equation}

We next need to calculate $L$.
The 2-magnon subspace of $\Hil_{\rm tot}$, consists of 6 basis vectors:
$$
\Psi_{jk}\, =\, S_j^- S_k^- \ket{\Uparrow} \qquad \text { for } \qquad 1\leq j<k\leq 4\, ,
$$
where $\ket{\Uparrow}$ is the all-up-spin vector.
We define two representative Ising basis vectors from $\Hil_{\rm tot}$, one for each translation orbit:
$$
\Psi_{12}\, =\,  \begin{minipage}{1.5cm}
\begin{tikzpicture}
\draw [] (0,0) circle (7.5mm);
\fill (45:7.5mm) circle (0.75mm);
\fill (315:7.5mm) circle (0.75mm);
\fill (225:7.5mm) circle (0.75mm);
\fill (135:7.5mm) circle (0.75mm);
\draw (45:1cm) node[] {$\downarrow$};
\draw (315:1cm) node[] {$\downarrow$};
\draw (225:1cm) node[] {$\uparrow$};
\draw (135:1cm) node[] {$\uparrow$};
\end{tikzpicture}
\end{minipage}\qquad \text { and } \quad
\Psi_{13}\, =\,  \begin{minipage}{1.5cm}
\begin{tikzpicture}
\draw [] (0,0) circle (7.5mm);
\fill (45:7.5mm) circle (0.75mm);
\fill (135:7.5mm) circle (0.75mm);
\fill (225:7.5mm) circle (0.75mm);
\fill (315:7.5mm) circle (0.75mm);
\draw (45:1cm) node[] {$\downarrow$};
\draw (135:1cm) node[] {$\uparrow$};
\draw (225:1cm) node[] {$\downarrow$};
\draw (315:1cm) node[] {$\uparrow$};
\end{tikzpicture}
\end{minipage}\qquad .
$$
Then, using the same symbol $T$ for translations in $\Hil_{\rm tot}$,
\begin{equation}
\label{eq:Lsecond}
\begin{split}
L \phi_{12,34}\, 
&=\, \Psi_{13} - \Psi_{23} - \Psi_{14} + \Psi_{24}\\
&=\, \Psi_{13} - T \Psi_{12} - T^3 \Psi_{12} + T \Psi_{13}\\
&=\, - (T + T^3) \Psi_{12} + (1 + T) \Psi_{13} \, .
\end{split}
\end{equation}
We define translation eigenvectors. But there are now two different sizes for the orbits.
\begin{itemize}
\item
For $z \in \{1,i,-1,-i\}$, we define
$$
\hat{\psi}_1(z)\, =\, \sum_{k=0}^3 z^k T^k \Psi_{12}\, .
$$
\item 
For $z \in \{1,-1\}$, we define
$$
\hat{\psi}_2(z)\, =\, \sum_{k=0}^1 z^k T^k \Psi_{13}\, .
$$
\end{itemize}
These are all linearly independent vectors in $\Hil_{\rm tot}$ for the cycle $C_4$.
From (\ref{eq:Lsecond}) we see that
$$
L \hat{\phi}(z)\, 
=\, -(z + z^3) \hat{\psi}_1(z) + (1+z^3) \hat{\psi}_2(z)\, ,
$$
for $z \in \{-1,1\}$.
This has trivial kernel.
Therefore, from (\ref{eq:Aspec2}) we may read off the spectrum for $H$ restricted to the 2 spin deviate subspace for $C_4$ 
$$
\operatorname{spec}(2H \restriction \Hil_{\rm tot}(0,0))\, =\, \{2,6\}\, .
$$
In particular,
$$
E_0(C_4,2)\, =\, 1\, .
$$
Comparing to (\ref{eq:E0first}) we see that we have the ``accidental degeneracy,''
$$
E_0(C_4,2)\, =\, E_0(C_4,1)\, .
$$
This satisfies both FOEL at level 1, because $E_0(C_4,1) \leq E_0(C_4,2)$,
and our ``opposite surmise,'' $E_0(C_{2n},n) \leq E_0(C_{2n},n-1)$, because of the equality.
This is the first example from our main results.

\subsubsection{Plots for $C_4$}
In Figure \ref{fig:first}, we have numerically diagonalized $2H$, using the Ising basis
in Matlab.
Matlab uses a standard implementation of the exact numerical diagonalization algorithm ``Lanczos,''
which is considered to be reliable for matrices of the sizes we considered.
We consider the square which is $C_N$ for $N=4$ in that figure.
All of the eigenvalues we have calculated here are displayed there, and the ``accidental degeneracy''
is circled.

\subsection{The hexagon $C_6$}

The Hubbard model on the hexagon was investigated by Heilmann and Lieb with a view towards
degeneracies \cite{HL}.
Since the Hubbard model is more complicated than the spin-$1/2$ Heisenberg model, this suggests that
the investigation of the Heisenberg model may also yield interesting results.
As a remark, we mention that they found many ``accidental'' degeneracies, i.e,
degeneracies which are not obviously related to a symmetry of the Hamiltonian.
It was this which
motivated the later work \cite{YHAS}.

Let us mention the easiest results.
We know that $E_0(C_6,0)=0$, because that is the ground state energy.
That is also the only eigenvalue of $H$ in the one-dimensional subspace
$\Hil_{\rm tot}(3,3)$ of 0-spin deviates.

The spectrum of $H$ in the 1-spin deviate subspace $\Hil_{\rm tot}(2,2)$
is also easy.
We may work in analogy with Subsection \ref{subsec:C4sd1}.
We obtain for $C_6$, that there are $A$ eigenvectors, $\hat{\phi}(z) \in V$
for $z \in \{1,e^{\pi i/3},e^{2\pi i/3},-1,e^{4\pi i/3},e^{5\pi i/3}\}$,
$$
\hat{\phi}(z)\, =\, \sum_{k=0}^{5} z^k T^k\qquad
\begin{minipage}{2.25cm}
\begin{tikzpicture}[xscale=1.5, yscale=1.5]
\draw [] (0,0) circle (7.5mm);
\fill (60:7.5mm) circle (0.75mm);
\fill (360:7.5mm) circle (0.75mm);
\fill (300:7.5mm) circle (0.75mm);
\fill (240:7.5mm) circle (0.75mm);
\fill (180:7.5mm) circle (0.75mm);
\fill (120:7.5mm) circle (0.75mm);
\draw[thick] (60:7.5mm) .. controls (60:4mm) and (360:4mm) .. (360:7.5mm);
\draw (360:4.2mm) [fill, rotate=30] +(0,0) -- +(-0.1,0.1) -- +(0.1,0.1);
\draw[] (300:7.5mm) -- (210:0mm);
\draw[] (240:7.5mm) -- (210:0mm);
\draw[] (180:7.5mm) -- (210:0mm);
\draw[] (120:7.5mm) -- (210:0mm);
\filldraw[thick,fill=white] (210:0mm) circle (1.5mm);
\draw (210:0mm) node[] {\small $\boldsymbol{\uparrow}$};
\end{tikzpicture}
\end{minipage}\qquad 
$$
and
$$
A \hat{\phi}(z)\, =\, (-2+z+z^5) \hat{\phi}(z)\, .
$$
However the kernel of the intertwining operator $L : V \to \Hil_{\rm tot}(2,2)$
consists of $\hat{\phi}(1)$.
Therefore, for $C_6$,
$$
\operatorname{spec}(2H \restriction \Hil_{\rm tot}(2,2))\, =\, \{1,3,4\}\, .
$$
More specifically,
$$
(2H - 1) L \hat{\phi}(e^{\pm \pi i/3})\, =\, (2H-3) L \hat{\phi}(e^{\pm 2 \pi i/3})\, =\,  (-2H-4) L \hat{\phi}(-1)\, =\, 0\, .
$$
In particular, $E_0(C_6,1) = 1/2$.
Another interesting fact relates to the energy eigenvalue $2H \hat{\phi}(-1)\, =\, 4 \hat{\phi}(-1)$.
As we will see, that eigenvalue of $2H$ also has an ``accidental degeneracy.''

\subsubsection{$C_6$, 2 spin deviate}
We now consider the case $s=1$.
This is the $2$ spin deviate subspace for $C_6$.

The diagrams we consider have 2 arcs.
We define three basic diagrams
$$
\phi_{14,23}\ =\quad 
\begin{minipage}{2.25cm}
\begin{tikzpicture}[xscale=1.5, yscale=1.5]
\draw [thick] (0,0) circle (7.5mm);
\fill (60:7.5mm) circle (0.75mm);
\fill (360:7.5mm) circle (0.75mm);
\fill (300:7.5mm) circle (0.75mm);
\fill (240:7.5mm) circle (0.75mm);
\fill (180:7.5mm) circle (0.75mm);
\fill (120:7.5mm) circle (0.75mm);
\draw[thick] (60:7.5mm) -- (240:7.5mm);
\draw[thick] (360:7.5mm) .. controls (360:4mm) and (300:4mm) .. (300:7.5mm);
\fill[rotate=-30] (0,0) -- (-0.1,0.1) -- (0.1,0.1);
\draw (360:4.2mm) [fill, rotate=-30] +(0,0) -- +(-0.1,0.1) -- +(0.1,0.1);
\draw[] (180:7.5mm) -- (150:3mm);
\draw[] (120:7.5mm) -- (150:3mm);
\filldraw[thick,fill=white] (150:3mm) circle (1.5mm);
\draw (150:3mm) node[] {\small $\boldsymbol{\uparrow}$};
\end{tikzpicture}
\end{minipage}\qquad ,\quad
\phi_{12,34}\ =\quad
\begin{minipage}{2.25cm}
\begin{tikzpicture}[xscale=1.5, yscale=1.5]
\draw [thick] (0,0) circle (7.5mm);
\fill (60:7.5mm) circle (0.75mm);
\fill (360:7.5mm) circle (0.75mm);
\fill (300:7.5mm) circle (0.75mm);
\fill (240:7.5mm) circle (0.75mm);
\fill (180:7.5mm) circle (0.75mm);
\fill (120:7.5mm) circle (0.75mm);
\draw[thick] (60:7.5mm) .. controls (60:4mm) and (360:4mm) .. (360:7.5mm);
\draw (360:4.2mm) [fill, rotate=30] +(0,0) -- +(-0.1,0.1) -- +(0.1,0.1);
\draw[thick] (300:7.5mm) .. controls (300:4mm) and (240:4mm) .. (240:7.5mm);
\draw (360:4.2mm) [fill, rotate=-90] +(0,0) -- +(-0.1,0.1) -- +(0.1,0.1);
\draw[] (180:7.5mm) -- (150:2mm);
\draw[] (120:7.5mm) -- (150:2mm);
\filldraw[thick,fill=white] (150:2mm) circle (1.5mm);
\draw (150:2mm) node[] {\small $\boldsymbol{\uparrow}$};
\end{tikzpicture}
\end{minipage}
$$
and
$$
\phi_{12,45}\ =\quad
\begin{minipage}{2.25cm}
\begin{tikzpicture}[xscale=1.5, yscale=1.5]
\draw [thick] (0,0) circle (7.5mm);
\fill (60:7.5mm) circle (0.75mm);
\fill (360:7.5mm) circle (0.75mm);
\fill (300:7.5mm) circle (0.75mm);
\fill (240:7.5mm) circle (0.75mm);
\fill (180:7.5mm) circle (0.75mm);
\fill (120:7.5mm) circle (0.75mm);
\draw[thick] (60:7.5mm) .. controls (60:4mm) and (360:4mm) .. (360:7.5mm);
\draw (360:4.2mm) [fill, rotate=30] +(0,0) -- +(-0.1,0.1) -- +(0.1,0.1);
\draw[thick] (240:7.5mm) .. controls (240:4mm) and (180:4mm) .. (180:7.5mm);
\draw (360:4.2mm) [fill, rotate=-150] +(0,0) -- +(-0.1,0.1) -- +(0.1,0.1);
\draw[] (300:7.5mm) -- (0,0);
\draw[] (120:7.5mm) -- (0,0);
\filldraw[thick,fill=white] (0,0) circle (1.5mm);
\draw (0,0) node[] {\small $\boldsymbol{\uparrow}$};
\end{tikzpicture}
\end{minipage}\qquad .
$$
With $T$ denoting the fundamental clockwise rotation, we may write
\begin{equation}
\label{eq:Asimple}
A
\begin{bmatrix}
\phi_{14,23} \\
\phi_{12,34} \\
\phi_{12,45}
\end{bmatrix}\,
=\,
\begin{bmatrix}
-2 & 2 + T + T^5 & 0 \\
1 & -4 & 1 + T ^2\\
0 & 1+T+T^3+T^4 & -4
\end{bmatrix}
\cdot
\begin{bmatrix}
\phi_{14,23} \\
\phi_{12,34} \\
\phi_{12,45}
\end{bmatrix}\, .
\end{equation}
We define translation eigenvectors based on these three translation orbit representatives.
The orbits have two different sizes.
\begin{itemize}
\item
For $z \in \{1,e^{\pi i/3},e^{2\pi i/3},-1,e^{4\pi i/3},e^{5\pi i/3}\}$, we define
$$
\hat{\phi}_1(z)\, =\, \sum_{k=0}^{5} z^k T^{k} \phi_{14,23}\qquad 
\text { and } \qquad 
\hat{\phi}_2(z)\, =\, \sum_{k=0}^{5} z^k T^k \phi_{12,34}\, .
$$
\item
For $z \in \{1,e^{2 \pi i/3},e^{4 \pi i/3}\}$, we define
$$
\hat{\phi}_3(z)\, =\, \sum_{k=0}^{2} z^k T^{k} \phi_{12,45}\, .
$$
\end{itemize}
Then (\ref{eq:Asimple}) may be used to deduce the action of $A$ in each translation eigenspace.
\begin{itemize}
\item 
For $z \in \{e^{\pi i /3},-1,e^{5\pi i/3}\}$, we have:
$$
\begin{bmatrix} x_1 & x_2 \end{bmatrix}
A 
\begin{bmatrix}
\hat{\phi}_1(z) \\
\hat{\phi}_2(z)
\end{bmatrix}\,
=\, 
\begin{bmatrix} x_1 & x_2 \end{bmatrix}
\begin{bmatrix} -2 & 2 + z + z^5 \\
1 & -4 \end{bmatrix}
\begin{bmatrix}
\hat{\phi}_1(z) \\
\hat{\phi}_2(z)
\end{bmatrix}\, .
$$
\item
For $z \in \{1,e^{2\pi i/3},e^{4\pi i /3}\}$,  the subgroup determined by the equation $z^3 = 1$,
$$
\begin{bmatrix} x_1 & x_2 & x_3 \end{bmatrix}
A 
\begin{bmatrix}
\hat{\phi}_1(z) \\
\hat{\phi}_2(z) \\
\hat{\phi}_3(z)
\end{bmatrix}\,
=\, 
\begin{bmatrix} x_1 & x_2 & x_3 \end{bmatrix}
\begin{bmatrix} -2 & 2 + z + z^2  & 0\\ 
1  & -4 & 2 + 2 z\\
0 & 1 + z^2 & -4\end{bmatrix}
\begin{bmatrix}
\hat{\phi}_1(z) \\
\hat{\phi}_2(z) \\
\hat{\phi}_3(z)
\end{bmatrix}\, .
$$
\end{itemize}
To diagonalize $A$ in each translation eigenspace (sector) one must solve for the left eigenvectors $[x_1,x_2]$ or $[x_1,x_2,x_3]$ of the matrices displayed above,
on the right-hand-sides of the equations.
From this we may deduce the spectrum of $A$
\begin{itemize}
\item
For $z \in \{e^{\pi i /3},-1,e^{5\pi i/3}\}$, which means $z^3=-1$, we have:
$$
\left[A - \left(-3 \pm \sqrt{3+z-z^2}\right)\right]
\left[\hat{\phi}_1(z)  
+ \left(-1\pm\sqrt{3+z-z^2}\right) \hat{\phi}_2(z)\right]\, 
=\, 0\, .
$$
\item
For $z \in \{e^{2\pi i /3},e^{4\pi i/3}\}$ we have:
\begin{gather*}
\left[A - (-3)\right] 
\left[\hat{\phi}_1(z) - \hat{\phi}_2(z) + 2 z^2 \hat{\phi}_3(z)\right]\,
=\, 0\, , \quad \text { and }\\
\left[ A - \left(\frac{-7 \pm \sqrt{17}}{2}\right)\right] 
\left[ \left(3 \pm \sqrt{17} \right) \hat{\phi}_1(z) 
+ 4 \hat{\phi}_2(z)
+ (1 \mp \sqrt{17}) \hat{\phi}_3(z) \right]\, =\, 0\, .
\end{gather*}
\item For $z = 1$, we have
\begin{gather*}
A \left[\hat{\phi}_1(z) + 2 \hat{\phi}_2(z) + 2 \hat{\phi}_3(z)\right]\, =\, 0 \quad \text { and }\\
\left[ A - \left(-5 \pm \sqrt{5}\right)\right] \left[\hat{\phi}_1(z) 
+ \left(-3 \pm \sqrt{5}\right)  \hat{\phi}_2(z) 
+ 2 \left(1 \mp \sqrt{5}\right) \hat{\phi}_3(z)\right]\, =\, 0\, .
\end{gather*}
\end{itemize}

Next we compute $L$ in this basis.
We define some basic Ising configuration vectors
$$
\Psi_{12}\, =\quad
\begin{minipage}{2.5cm}
\begin{tikzpicture}[xscale=1, yscale=1]
\draw [] (0,0) circle (7.5mm);
\fill (60:7.5mm) circle (0.75mm);
\fill (360:7.5mm) circle (0.75mm);
\fill (300:7.5mm) circle (0.75mm);
\fill (240:7.5mm) circle (0.75mm);
\fill (180:7.5mm) circle (0.75mm);
\fill (120:7.5mm) circle (0.75mm);
\draw (60:1cm) node {$\downarrow$};
\draw (360:1cm) node {$\downarrow$};
\draw (300:1cm) node {$\uparrow$};
\draw (240:1cm) node {$\uparrow$};
\draw (180:1cm) node {$\uparrow$};
\draw (120:1cm) node {$\uparrow$};
\end{tikzpicture}
\end{minipage}\quad\, ,\qquad
\Psi_{13}\, =\quad
\begin{minipage}{2.5cm}
\begin{tikzpicture}[xscale=1, yscale=1]
\draw [] (0,0) circle (7.5mm);
\fill (60:7.5mm) circle (0.75mm);
\fill (360:7.5mm) circle (0.75mm);
\fill (300:7.5mm) circle (0.75mm);
\fill (240:7.5mm) circle (0.75mm);
\fill (180:7.5mm) circle (0.75mm);
\fill (120:7.5mm) circle (0.75mm);
\draw (60:1cm) node {$\downarrow$};
\draw (360:1cm) node {$\uparrow$};
\draw (300:1cm) node {$\downarrow$};
\draw (240:1cm) node {$\uparrow$};
\draw (180:1cm) node {$\uparrow$};
\draw (120:1cm) node {$\uparrow$};
\end{tikzpicture}
\end{minipage}
$$
and
$$
\Psi_{14}\, =\quad
\begin{minipage}{2.5cm}
\begin{tikzpicture}[xscale=1, yscale=1]
\draw [] (0,0) circle (7.5mm);
\fill (60:7.5mm) circle (0.75mm);
\fill (360:7.5mm) circle (0.75mm);
\fill (300:7.5mm) circle (0.75mm);
\fill (240:7.5mm) circle (0.75mm);
\fill (180:7.5mm) circle (0.75mm);
\fill (120:7.5mm) circle (0.75mm);
\draw (60:1cm) node {$\downarrow$};
\draw (360:1cm) node {$\uparrow$};
\draw (300:1cm) node {$\uparrow$};
\draw (240:1cm) node {$\downarrow$};
\draw (180:1cm) node {$\uparrow$};
\draw (120:1cm) node {$\uparrow$};
\end{tikzpicture}
\end{minipage}\quad .
$$
Using $T$ for the usual shift, one may calculate the action of the intertwining operator $L : V \to \Hil_{\rm tot}(1,1)$ for $C_6$,
\begin{equation}
\label{eq:Lsimple}
L
\begin{bmatrix}
\phi_{14,23} \\
\phi_{12,34} \\
\phi_{12,45}
\end{bmatrix}\,
=\,
\begin{bmatrix}
1+T^2 & -(1+T) & 0 \\
-T & 1+T & -1\\
0 & -(T+T^4) & 1+T
\end{bmatrix}
\cdot
\begin{bmatrix}
\Psi_{12} \\
\Psi_{13} \\
\Psi_{14}
\end{bmatrix}\, .
\end{equation}
We define translation eigenvectors, as follows.
\begin{itemize}
\item
For $z \in \{1,e^{\pi i/3},e^{2\pi i/3},-1,e^{4\pi i/3},e^{5\pi i/3}\}$, we define
$$
\hat{\psi}_1(z)\, =\, \sum_{k=0}^{5} z^k T^{k} \Psi_{12} \qquad \text { and } \qquad 
\hat{\psi}_2(z)\, =\, \sum_{k=0}^{5} z^k T^{k} \Psi_{13}\, .
$$
\item
For $z \in \{1,e^{2\pi i/3},e^{4\pi i/3}\}$, we define
$$
\hat{\psi}_3(z)\, =\, \sum_{k=0}^{2} z^k T^{k} \Psi_{14}\, .
$$
\end{itemize}
Then, from (\ref{eq:Lsimple}), we have the following formulas for $L$ in each translation eigenspace.
\begin{itemize}
\item
For $z \in \{e^{\pi i /3},-1,e^{5\pi i/3}\}$, which means $z^3=-1$, 
$$
\begin{bmatrix} x_1 & x_2 \end{bmatrix}
L \begin{bmatrix}
\hat{\phi}_1(z) \\
\hat{\phi}_2(z)
\end{bmatrix}\, 
=\, 
\begin{bmatrix} x_1 & x_2 \end{bmatrix}
\begin{bmatrix} 1-z & -1+z^2 \\
z^2 & 1-z^2 \end{bmatrix}
\begin{bmatrix}
\hat{\phi}_1(z) \\
\hat{\phi}_2(z)
\end{bmatrix}\,  .
$$
\item
For $z \in \{1,e^{2\pi i/3},e^{4\pi i /3}\}$, which means $z^3=1$, we have
$$
\begin{bmatrix} x_1 & x_2 & x_3 \end{bmatrix}
L
\begin{bmatrix}
\hat{\phi}_1(z) \\
\hat{\phi}_2(z) \\
\hat{\phi}_3(z) 
\end{bmatrix}\,
=\, 
\begin{bmatrix} x_1 & x_2 & x_3 \end{bmatrix}
\begin{bmatrix}  
1+z & -1-z^2 & 0\\ 
-z^2 & 1+z^2 & -2\\
0 & -z^2 & 1+z^2\end{bmatrix}
\cdot
\begin{bmatrix}
\hat{\psi}_1(z) \\
\hat{\psi}_2(z) \\
\hat{\psi}_3(z)
\end{bmatrix}\, .
$$
\end{itemize}
From this, we can see that the kernel of $L$ is spanned by the vectors:
\begin{gather*}
\hat{\phi}_1(1) + 2 \hat{\phi}_2(1) + 2 \hat{\phi}_3(1)\, ,\
\hat{\phi}_1(-1) - 2 \hat{\phi}_2(-1)\, ,\
\hat{\phi}_1(e^{\pm \pi i/3}) + \hat{\phi}_2(e^{\pm \pi i/3}) \quad \text { and }\\
\hat{\phi}_1(e^{\pm 2 \pi i/3}) - \hat{\phi}_2(e^{\pm 2 \pi i/3}) + 2 e^{\pm 4 \pi i/3} \hat{\phi}_3(e^{\pm 2 \pi i/3})\, .
\end{gather*}
Each of these vectors is an eigenvector of $A$. 
The eigenvalues associated to these eigenvectors, which need to be removed are $0$, $-1$, $-3$, $-4$.
This leaves the following eigenvalues of $H$ in the 2 spin deviate sector for $C_6$:
$$
\sigma(2H \restriction \Hil_{\rm tot}(1,1))\, 
=\, \left\{\frac{7-\sqrt{17}}{2}\, ,\ 2\, ,\ 5-\sqrt{5}\, ,\ 5\, ,\ 5+\sqrt{5}\, ,\ \frac{7+\sqrt{17}}{2}\right\}\, .
$$
In particular,
$$
E_0(C_6,2)\, =\, \frac{7-\sqrt{17}}{4}\, .
$$

\subsubsection{$C_6$, 3 spin deviate}
We now consider singlets.
Graphically, this corresponds to total matchings.
We consider two representatives of translation orbits:
$$
\phi_{14,23,56}\, =\quad 
\begin{minipage}{2.25cm}
\begin{tikzpicture}[xscale=1.5, yscale=1.5]
\draw [thick] (0,0) circle (7.5mm);
\fill (60:7.5mm) circle (0.75mm);
\fill (360:7.5mm) circle (0.75mm);
\fill (300:7.5mm) circle (0.75mm);
\fill (240:7.5mm) circle (0.75mm);
\fill (180:7.5mm) circle (0.75mm);
\fill (120:7.5mm) circle (0.75mm);
\draw[thick] (60:7.5mm) -- (240:7.5mm);
\draw[thick] (360:7.5mm) .. controls (360:4mm) and (300:4mm) .. (300:7.5mm);
\fill[rotate=-30] (0,0) -- (-0.1,0.1) -- (0.1,0.1);
\draw (360:4.2mm) [fill, rotate=-30] +(0,0) -- +(-0.1,0.1) -- +(0.1,0.1);
\draw[thick] (180:7.5mm) .. controls (180:4mm) and (120:4mm) .. (120:7.5mm);
\draw (360:4.2mm) [fill, rotate=150] +(0,0) -- +(-0.1,0.1) -- +(0.1,0.1);
\end{tikzpicture}
\end{minipage}\qquad 
\text { and } \qquad
\phi_{12,34,56}\, =\quad
\quad 
\begin{minipage}{2.25cm}
\begin{tikzpicture}[xscale=1.5, yscale=1.5]
\draw [thick] (0,0) circle (7.5mm);
\fill (60:7.5mm) circle (0.75mm);
\fill (360:7.5mm) circle (0.75mm);
\fill (300:7.5mm) circle (0.75mm);
\fill (240:7.5mm) circle (0.75mm);
\fill (180:7.5mm) circle (0.75mm);
\fill (120:7.5mm) circle (0.75mm);
\draw[thick] (60:7.5mm) .. controls (60:4mm) and (360:4mm) .. (360:7.5mm);
\draw (360:4.2mm) [fill, rotate=30] +(0,0) -- +(-0.1,0.1) -- +(0.1,0.1);
\draw[thick] (300:7.5mm) .. controls (300:4mm) and (240:4mm) .. (240:7.5mm);
\draw (360:4.2mm) [fill, rotate=-90] +(0,0) -- +(-0.1,0.1) -- +(0.1,0.1);
\draw[thick] (180:7.5mm) .. controls (180:4mm) and (120:4mm) .. (120:7.5mm);
\draw (360:4.2mm) [fill, rotate=150] +(0,0) -- +(-0.1,0.1) -- +(0.1,0.1);
\end{tikzpicture}
\end{minipage}\quad .
$$
Note that for $\phi_{14,23,56}$, rotation by $\pi$ changes the orientation of the arc connecting $1$ and $4$.
In principle, we should also introduce the basis vector $\phi_{21,43,56}$ which is related to 
$\phi_{12,34,56}$ by reversing all orientations.
However, we find that the translation on $\phi_{12,34,56}$ does not reverse orientation.
Moreover, we are not trying to exploit a ``good signs'' condition, using the Perron-Frobenius theorem.
Therefore, we merely allow for the possibility of taking $-\phi_{12,34,56}$
instead of $\phi_{12,34,56}$.

Letting $T$ be the usual translation, 
$$
A \begin{bmatrix} \phi_{14,23,56} \\ \phi_{12,34,56} \end{bmatrix}\,
=\, \begin{bmatrix} -4 & 2-2T \\ 1+T^2+T^4 & -6 \end{bmatrix}
\begin{bmatrix} \phi_{14,23,56} \\ \phi_{12,34,56} \end{bmatrix}\, .
$$
Note the generalized matrix entry $2-2T$, rather than $2+2T$.
This ``bad sign'' in an off-diagonal entry is precisely because we did not also include the vector
$\phi_{21,43,56}$ which reverses all three orientations from $\phi_{12,34,56}$, but rather simply
took this to be $-\phi_{12,34,56}$.

We define translation eigenvectors.
\begin{itemize}
\item For $z \in \{1,e^{\pi i/3},e^{2\pi i/3},-1,e^{4\pi i/3},e^{5\pi i/3}\}$, we define
$$
\hat{\phi}_1(z)\, =\, \sum_{k=0}^{5} z^k T^{k} \phi_{14,23,56}\, .
$$
\item For $z \in \{1,-1\}$, we define
$$
\hat{\phi}_2(z)\, =\, \sum_{k=0}^{1} z^k T^k \phi_{12,34,56}\, .
$$
\end{itemize}
Then we have the following action of $A$.
\begin{itemize}
\item For $z \in \{e^{\pi i/3},e^{2\pi i/3},e^{4\pi i/3},e^{5 \pi i/3}\}$,
$$
A \hat{\phi}_1(z)\, =\, -4 \hat{\phi}_1(z)\, .
$$
\item For $z \in \{1,-1\}$,
$$
\begin{bmatrix} x_1 & x_2 \end{bmatrix}
A \begin{bmatrix} \hat{\phi}_1(z) \\ \hat{\phi}_2(z) \end{bmatrix}\,
=\, \begin{bmatrix} x_1 & x_2 \end{bmatrix}
\begin{bmatrix} - 4 & 2(1-z)(1+z^2+z^4) \\ 1 & -6 \end{bmatrix}
\begin{bmatrix} \hat{\phi}_1(z) \\ \hat{\phi}_2(z) \end{bmatrix}\, .
$$
\end{itemize}
From this we deduce the spectrum of $A$, as follows.
\begin{itemize}
\item For $z \in \{e^{\pi i/3},e^{2\pi i/3},e^{4\pi i/3},e^{5 \pi i/3}\}$,
$$
\left[A - (-4)\right] \hat{\phi}_1(z)\, =\, 0\, .
$$
\item For $z = -1$,
$$
\left[A-\left(-5 \pm \sqrt{13}\right)\right] \left[\hat{\phi}_1(z) + \left(-1 \pm \sqrt{13}\right) \hat{\phi}_2(z)\right]\, =\, 0\, .
$$
\item For $z=1$,
$$
\left[A-(-4)\right] \hat{\phi}_1(z)\, =\, 0 \quad \text { and } \quad 
\left[A - (-6)\right] \left[\hat{\phi}_1(z) -2 \hat{\phi}_2(z)\right]\, =\, 0\, .
$$
\end{itemize}
Next we calculate the action of $L$.
We define the Ising basis vectors
\begin{gather*}
\Psi_{123}\, =\quad
\begin{minipage}{2.25cm}
\begin{tikzpicture}[xscale=1, yscale=1]
\draw [] (0,0) circle (7.5mm);
\fill (60:7.5mm) circle (0.75mm);
\fill (360:7.5mm) circle (0.75mm);
\fill (300:7.5mm) circle (0.75mm);
\fill (240:7.5mm) circle (0.75mm);
\fill (180:7.5mm) circle (0.75mm);
\fill (120:7.5mm) circle (0.75mm);
\draw (60:1cm) node {$\downarrow$};
\draw (360:1cm) node {$\downarrow$};
\draw (300:1cm) node {$\downarrow$};
\draw (240:1cm) node {$\uparrow$};
\draw (180:1cm) node {$\uparrow$};
\draw (120:1cm) node {$\uparrow$};
\end{tikzpicture}
\end{minipage}\quad ,\qquad 
\Psi_{124}\, =\quad 
\begin{minipage}{2.25cm}
\begin{tikzpicture}[xscale=1, yscale=1]
\draw [] (0,0) circle (7.5mm);
\fill (60:7.5mm) circle (0.75mm);
\fill (360:7.5mm) circle (0.75mm);
\fill (300:7.5mm) circle (0.75mm);
\fill (240:7.5mm) circle (0.75mm);
\fill (180:7.5mm) circle (0.75mm);
\fill (120:7.5mm) circle (0.75mm);
\draw (60:1cm) node {$\downarrow$};
\draw (360:1cm) node {$\downarrow$};
\draw (300:1cm) node {$\uparrow$};
\draw (240:1cm) node {$\downarrow$};
\draw (180:1cm) node {$\uparrow$};
\draw (120:1cm) node {$\uparrow$};
\end{tikzpicture}
\end{minipage}\quad ,\\[10pt]
\Psi_{125}\, =\quad
\begin{minipage}{2.25cm}
\begin{tikzpicture}[xscale=1, yscale=1]
\draw [] (0,0) circle (7.5mm);
\fill (60:7.5mm) circle (0.75mm);
\fill (360:7.5mm) circle (0.75mm);
\fill (300:7.5mm) circle (0.75mm);
\fill (240:7.5mm) circle (0.75mm);
\fill (180:7.5mm) circle (0.75mm);
\fill (120:7.5mm) circle (0.75mm);
\draw (60:1cm) node {$\downarrow$};
\draw (360:1cm) node {$\downarrow$};
\draw (300:1cm) node {$\uparrow$};
\draw (240:1cm) node {$\uparrow$};
\draw (180:1cm) node {$\downarrow$};
\draw (120:1cm) node {$\uparrow$};
\end{tikzpicture}
\end{minipage}\qquad \text { and } \qquad
\Psi_{135}\, =\quad 
\begin{minipage}{2.25cm}
\begin{tikzpicture}[xscale=1, yscale=1]
\draw [] (0,0) circle (7.5mm);
\fill (60:7.5mm) circle (0.75mm);
\fill (360:7.5mm) circle (0.75mm);
\fill (300:7.5mm) circle (0.75mm);
\fill (240:7.5mm) circle (0.75mm);
\fill (180:7.5mm) circle (0.75mm);
\fill (120:7.5mm) circle (0.75mm);
\draw (60:1cm) node {$\downarrow$};
\draw (360:1cm) node {$\uparrow$};
\draw (300:1cm) node {$\downarrow$};
\draw (240:1cm) node {$\uparrow$};
\draw (180:1cm) node {$\downarrow$};
\draw (120:1cm) node {$\uparrow$};
\end{tikzpicture}
\end{minipage}\quad .
\end{gather*}
Then
$$
L 
\begin{bmatrix} \phi_{14,23,56} \\ \phi_{12,34,56} \end{bmatrix}\,
=\, \begin{bmatrix} T^2 - T^5 & -T^2+T^5 & 1-T^3 & -1 + T \\
0 & -T-T^3-T^5 & T+T^3+T^5 & 1-T 
\end{bmatrix}
\begin{bmatrix} \Psi_{123} \\ \Psi_{124} \\ \Psi_{125} \\ \Psi_{135} \end{bmatrix}\, .
$$
We define translation eigenvectors as follows.
\begin{itemize}
\item For $z \in \{1,e^{\pi i/3},e^{2\pi i/3},-1,e^{4\pi i/3},e^{5\pi i/3}\}$, we define
$$
\begin{bmatrix}
\hat{\psi}_1(z)\\
\hat{\psi}_2(z)\\
\hat{\psi}_3(z)
\end{bmatrix}\, =\, \sum_{k=0}^{5} z^k T^{k} \begin{bmatrix}\Psi_{123}\\ \Psi_{124} \\ \Psi_{125} \end{bmatrix}\, .
$$
\item For $z \in \{1,-1\}$, we define
$$
\hat{\psi}_4(z)\, =\, \sum_{k=0}^{1} z^k T^k \Psi_{135}\, .
$$
\end{itemize}
Then we may rewrite the action of $L$ as follows.
\begin{itemize}
\item For $z \in \{e^{\pi i/3},e^{2\pi i/3},e^{4\pi i/3},e^{5 \pi i/3}\}$,
$$
L \hat{\phi}_1(z)\, =\, (1-z^3)\begin{bmatrix} -z & z & 1 \end{bmatrix} \begin{bmatrix}
\hat{\psi}_1(z)\\
\hat{\psi}_2(z)\\
\hat{\psi}_3(z)
\end{bmatrix}\, .
$$
\item For $z \in \{1,-1\}$, 
$$
\begin{bmatrix} x_1 & x_2 \end{bmatrix} L \begin{bmatrix} \phi_1(z) \\ \phi_2(z) \end{bmatrix}\,
=\, \begin{bmatrix} (1-z) x_1 & x_2 \end{bmatrix}
\begin{bmatrix} -z & z & 1 & -3\\
0 & -z & z & 1-z\end{bmatrix} 
\begin{bmatrix}
\hat{\psi}_1(z)\\
\hat{\psi}_2(z)\\
\hat{\psi}_3(z)\\
\hat{\psi}_4(z)
\end{bmatrix}\, .
$$
\end{itemize}
From this, we may deduce that the kernel of $L$ is spanned by the vectors
$$
\hat{\phi}(1) \quad \text { and } \quad \hat{\phi}(e^{\pm 2 \pi i /3})\, .
$$
Removing these vectors as eigenvectors of $A$ leaves the spectrum of $H$
in the 3 spin deviate space for $C_6$:
$$
\operatorname{spec}(2H \restriction \Hil_{\rm tot}(0,0))\, =\, \left \{ 5-\sqrt{13}\, ,\ 4\, ,\ 6\, ,\ 5+\sqrt{13}\right \}\, .
$$
In particular this means that
$$
E_0(C_6,3)\, =\, \frac{5-\sqrt{13}}{2}\, .
$$
Comparing, we see that
$$
E_0(C_6,3)\, =\, \frac{5-\sqrt{13}}{2}\, =\, 0.697224362\, <\,
E_0(C_6,2)\, =\, \frac{7-\sqrt{17}}{4}\, =\, 0.719223593\, .
$$
So this gives a violation of FOEL at level $2$ since $n=6$ implies that spin $s=1$
is $(n/2) - 2$.
This is the second example of our main results.

\subsubsection{Plots for $C_6$}
In Figure \ref{fig:first}, we have plotted the results of having numerically diagonalized $2H$, 
on the ring $C_N$ for several values of $N$.
The hexagon is $N=6$.
All the eigenvalues we mentioned are represented there.
In addition to the violation of FOEL, there is also an ``accidental degeneracy.''
The eigenvalue $4$ occurs both in the 1 spin deviate subspace $\Hil_{\rm tot}(2,2)$ 
and the 3 spin deviate subspace $\Hil_{\rm tot}(0,0)$.
The violation of FOEL is indicated with an arrow, and the ``accidental degeneracy''
is circled.

\section{Counterexamples through exact Lanczos diagonalization}

\begin{figure}
\begin{center}
\begin{tikzpicture}
\draw (0,0) node[above] {\includegraphics[height=8cm]{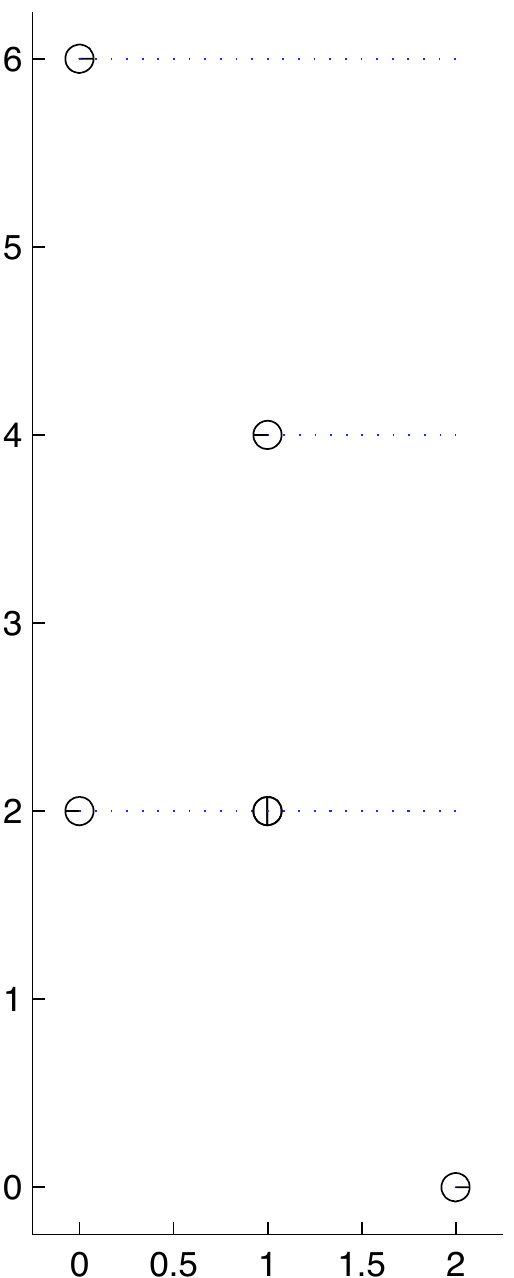}};
\draw (0,0) node[below] {$N=4$};
\draw (4,0) node[above] {\includegraphics[height=8cm]{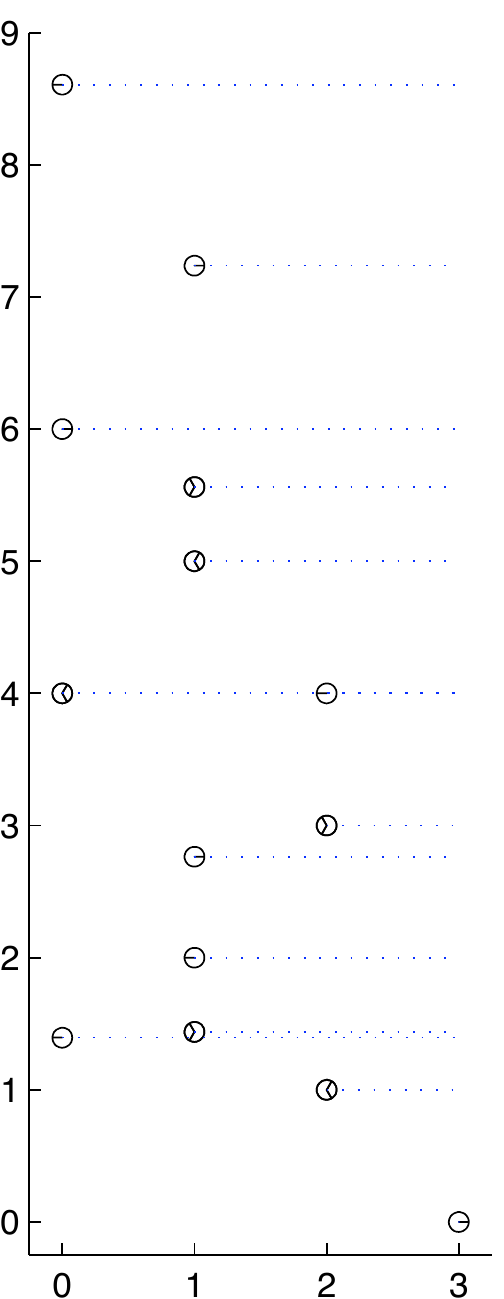}};
\draw (4,0) node[below] {$N=6$};
\draw (8,0) node[above] {\includegraphics[height=8cm]{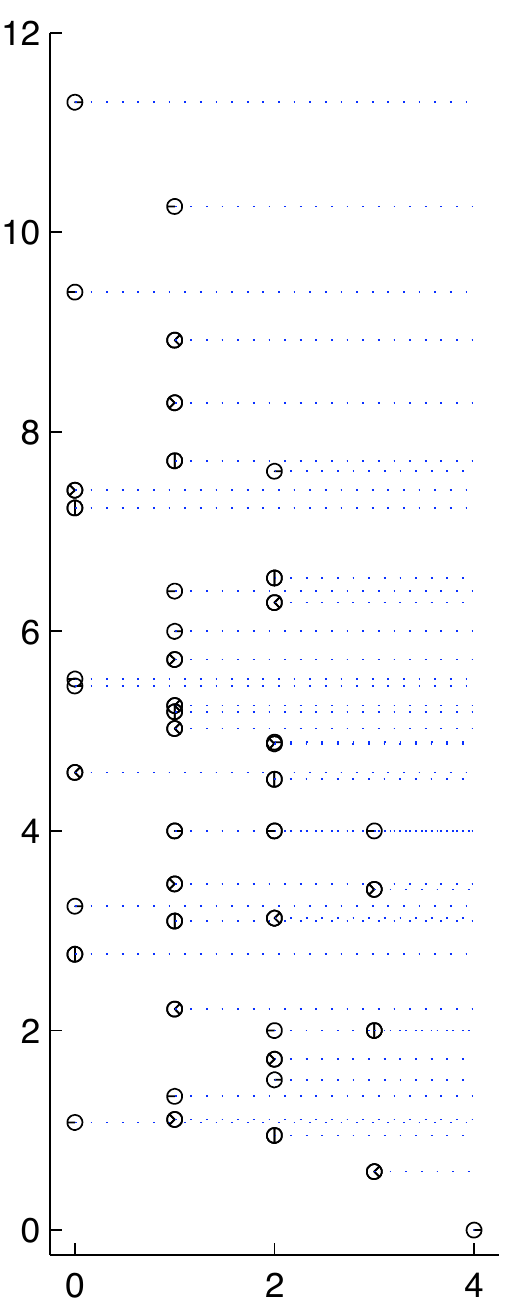}};
\draw (8,0) node[below] {$N=8$};
\draw[thick,red,dashed] (-0.5,3) ellipse (1cm and .275cm);
\draw[thick, red,dashed,xshift=5.05cm,yshift=1cm] (-1.4,2.85) ellipse (1cm and .285cm);
\draw[thick, red,dashed,->] (3.6,1.95) -- (2.85,1.9);
\draw[thick, red,dashed,->] (7.5,1.1) -- (6.85,1);
\draw[thick, red,dashed] (8.1,3) ellipse (0.85cm and .2cm);
\draw[thick, red,dashed] (8.475,1.77) ellipse (0.55cm and .1cm);
\draw (2,-9) node[above] {\includegraphics[height=8cm]{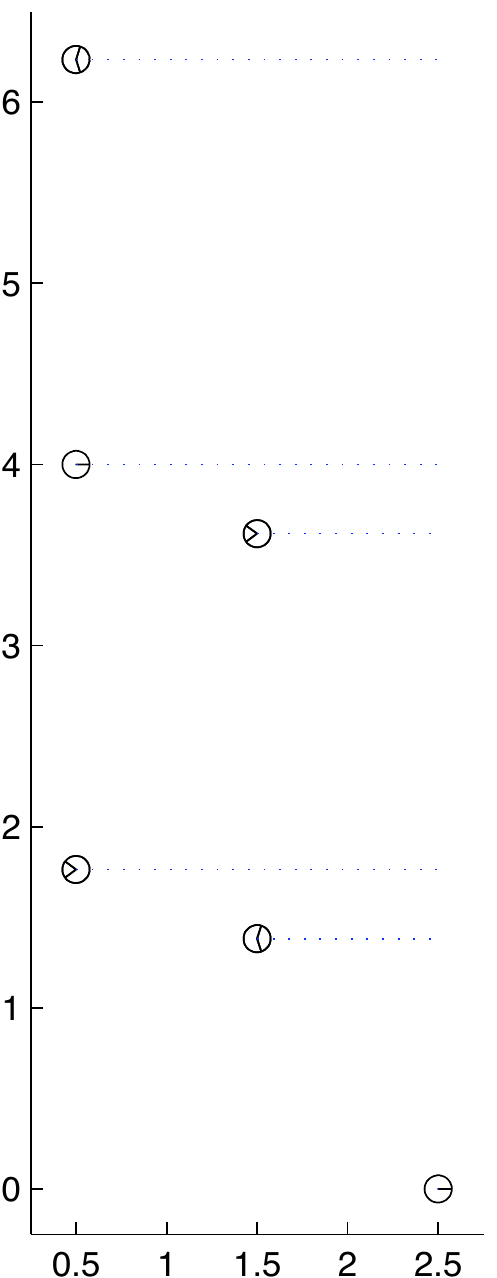}};
\draw (2,-9) node[below] {$N=5$};
\draw (6,-9) node[above] {\includegraphics[height=8cm]{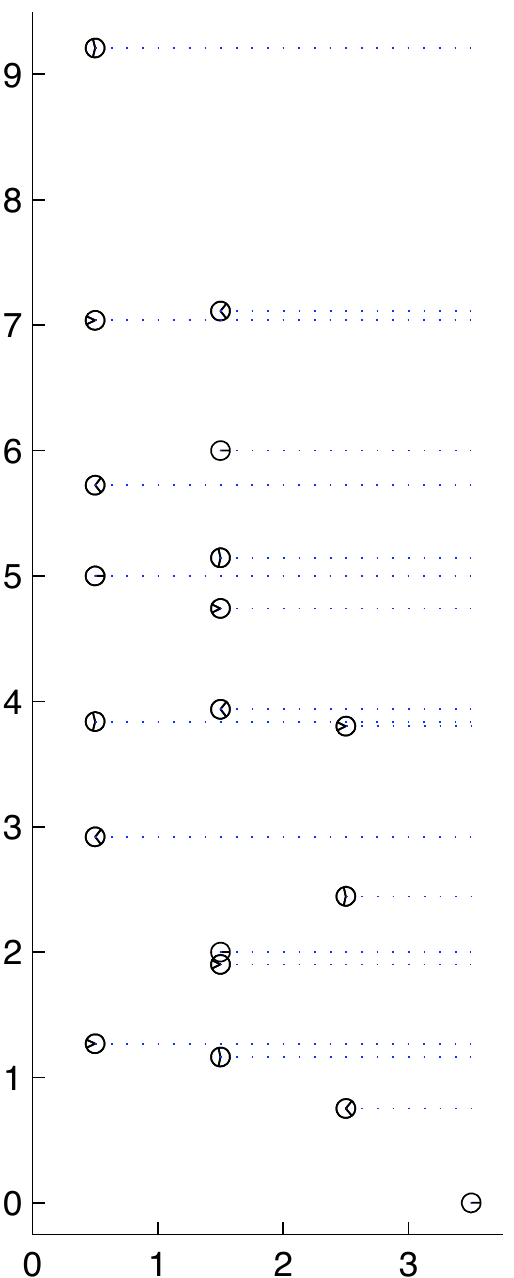}};
\draw (6,-9) node[below] {$N=7$};
\end{tikzpicture}
\end{center}
\caption{\label{fig:first} (Color online.) We have plotted $\operatorname{spec}(2H \restriction \Hil_{\rm tot}(s,s))$ for the cycles $C_{N}$
for $N=4,5,\dots,8$. The horizontal axis is $s$ for $s \in \{(L/2)-k\, :\, k = 0,\dots,n\}$ where we define $n=\lfloor N/2 \rfloor$. 
Each plot point is a circle with a sector inscribed, which corresponds to the translation eigenvalue(s) of the associated eigenvector(s).
The dashed circles represent ``accidental degeneracies,'' while the arrows indicate violations of FOEL.}
\end{figure}
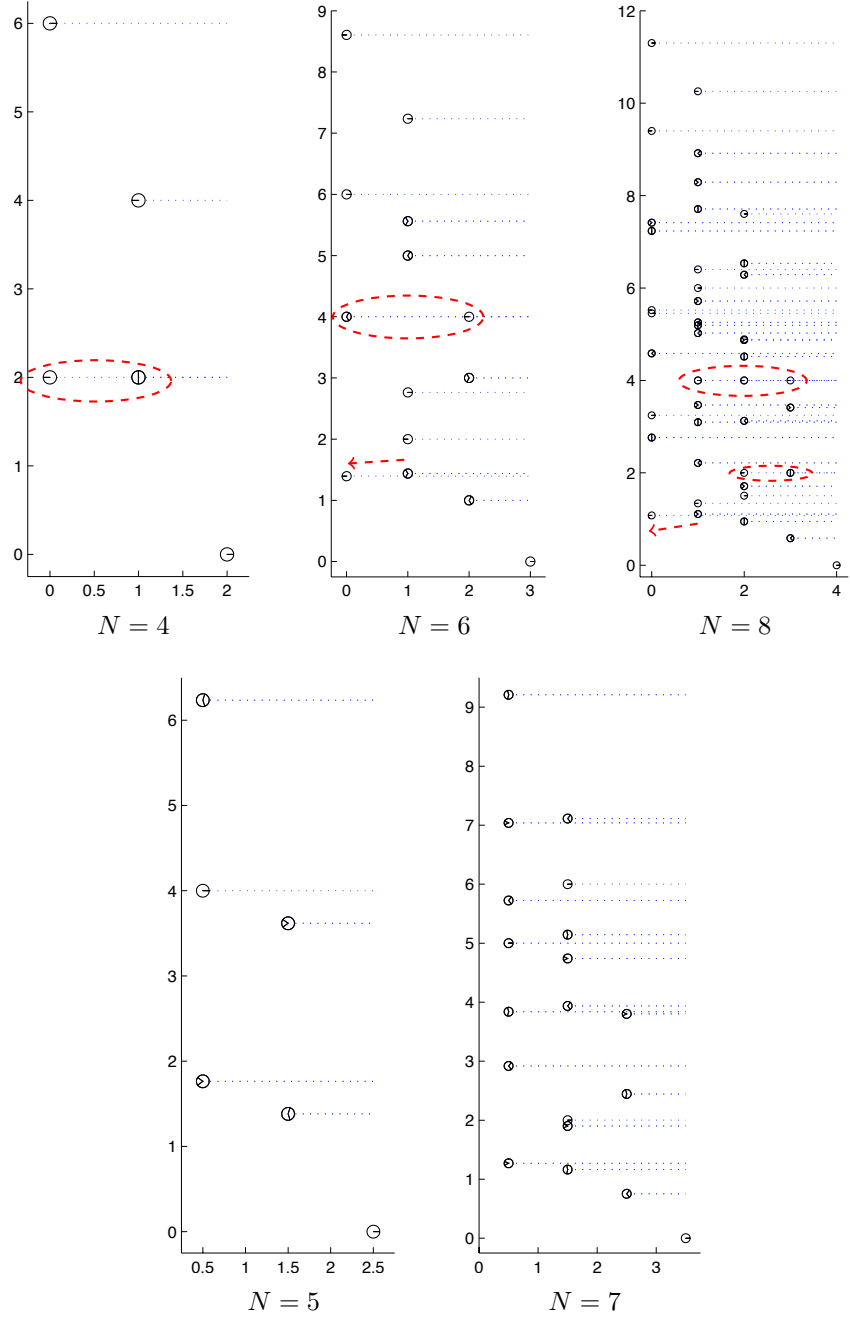

\begin{figure}
\begin{center}
\begin{tikzpicture}
\draw (0,5) node[above] {\includegraphics[width=8cm]{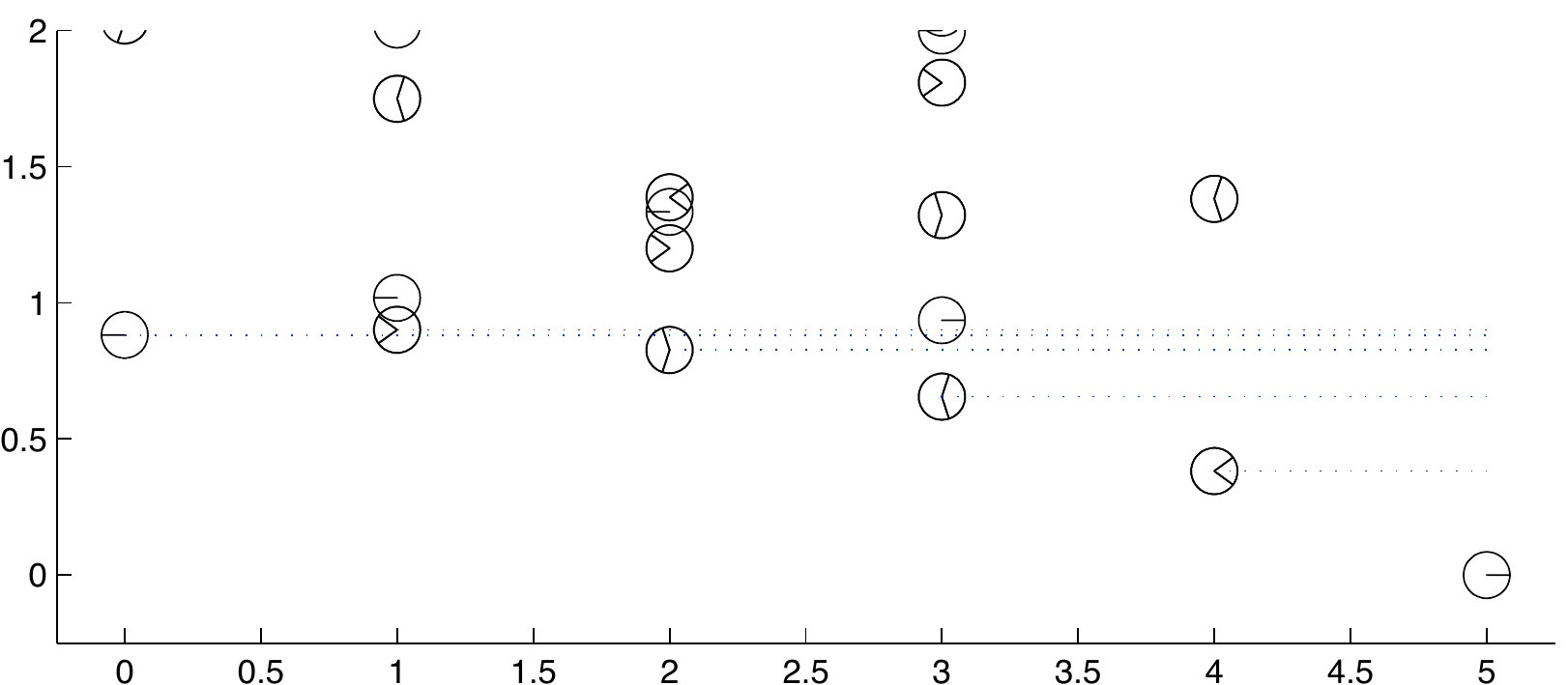}};
\draw (0,5) node[below] {$N=10$};
\draw (0,0) node[above] {\includegraphics[width=10cm]{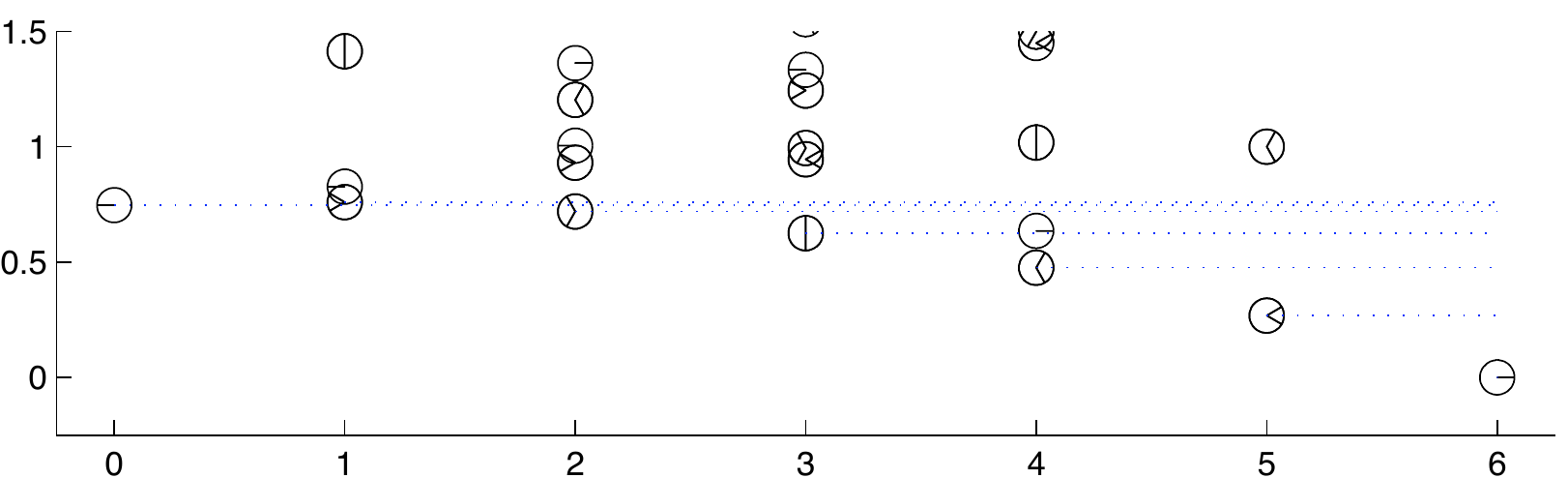}};
\draw (0,0) node[below] {$N=12$};
\draw (0,-5) node[above] {\includegraphics[width=12cm]{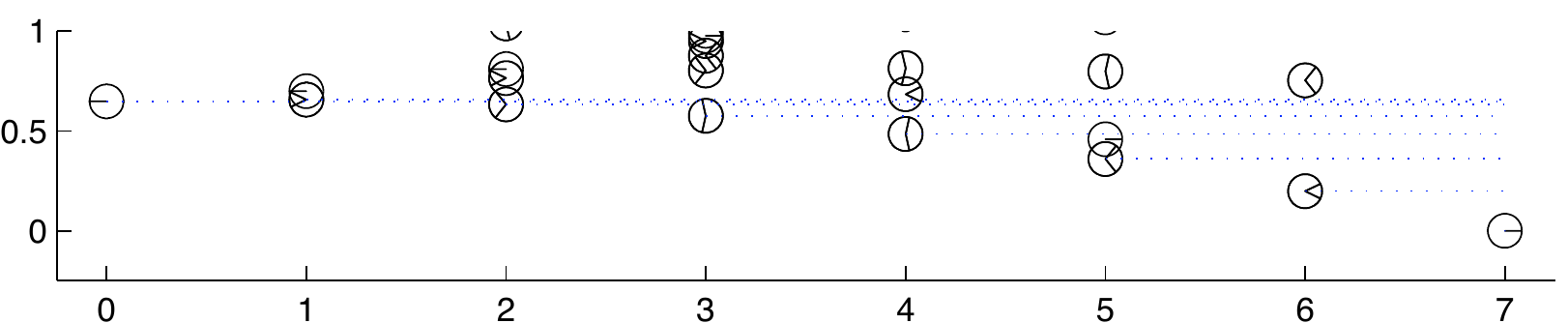}};
\draw (0,-5) node[below] {$N=14$};
\end{tikzpicture}
\end{center}
\caption{\label{fig:second} (Color online.) We have plotted $\operatorname{spec}(2H \restriction \Hil_{\rm tot}(s,s))$ for the cycles $C_{N}$
for $N=10,12,14$. The horizontal axis is $s$ for $s \in \{n-k\, :\, k = 0,\dots n\}$ where we define $n=N/2$. 
Each plot point is a circle with a sector inscribed, which corresponds to the translation eigenvalue(s) of the associated eigenvector(s).
In order to compare eigenvalues, we have put a dotted line emanating from each eigenvalue.
By zooming in, one can see that FOEL is violated, as claimed in the main results, for these examples.}
\end{figure}
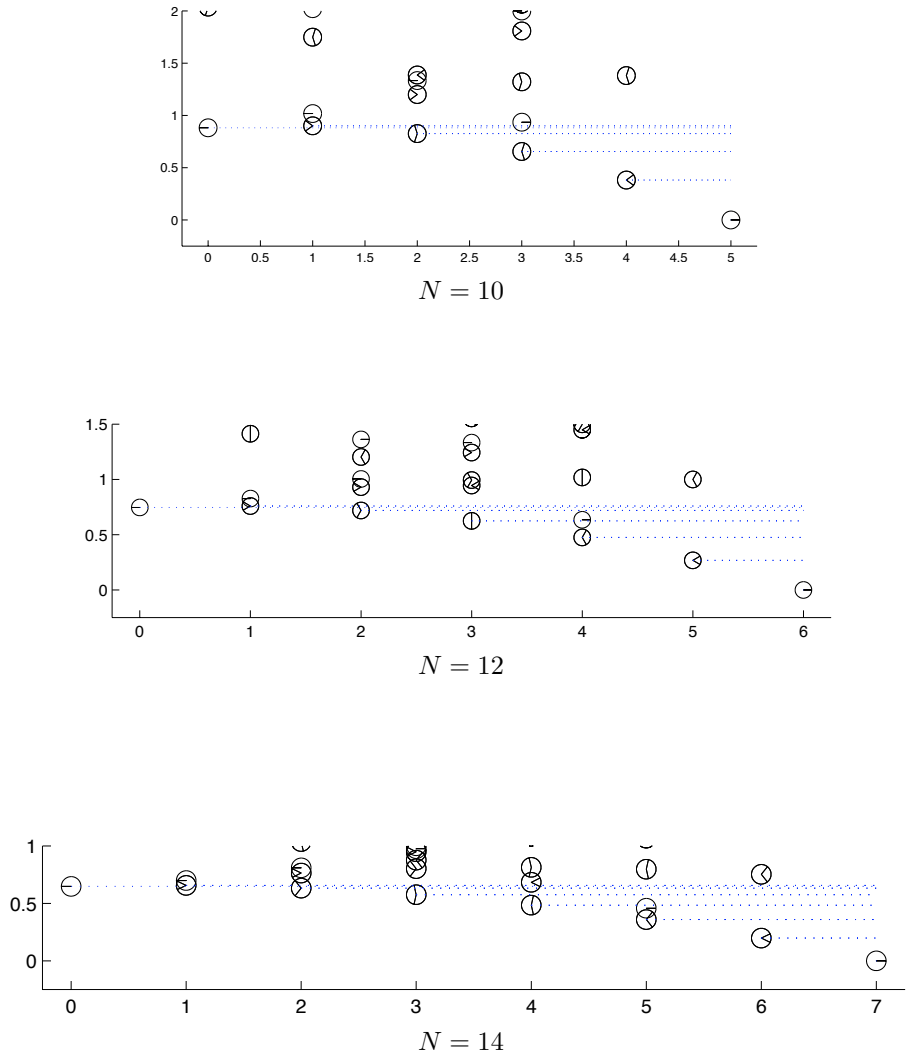

\begin{figure}
\begin{center}
\begin{tikzpicture}
%\draw (0,5) node[above] {\includegraphics[width=12cm]{L15grey.pdf}};
\draw (0,5) node[above] {\includegraphics[width=12cm]{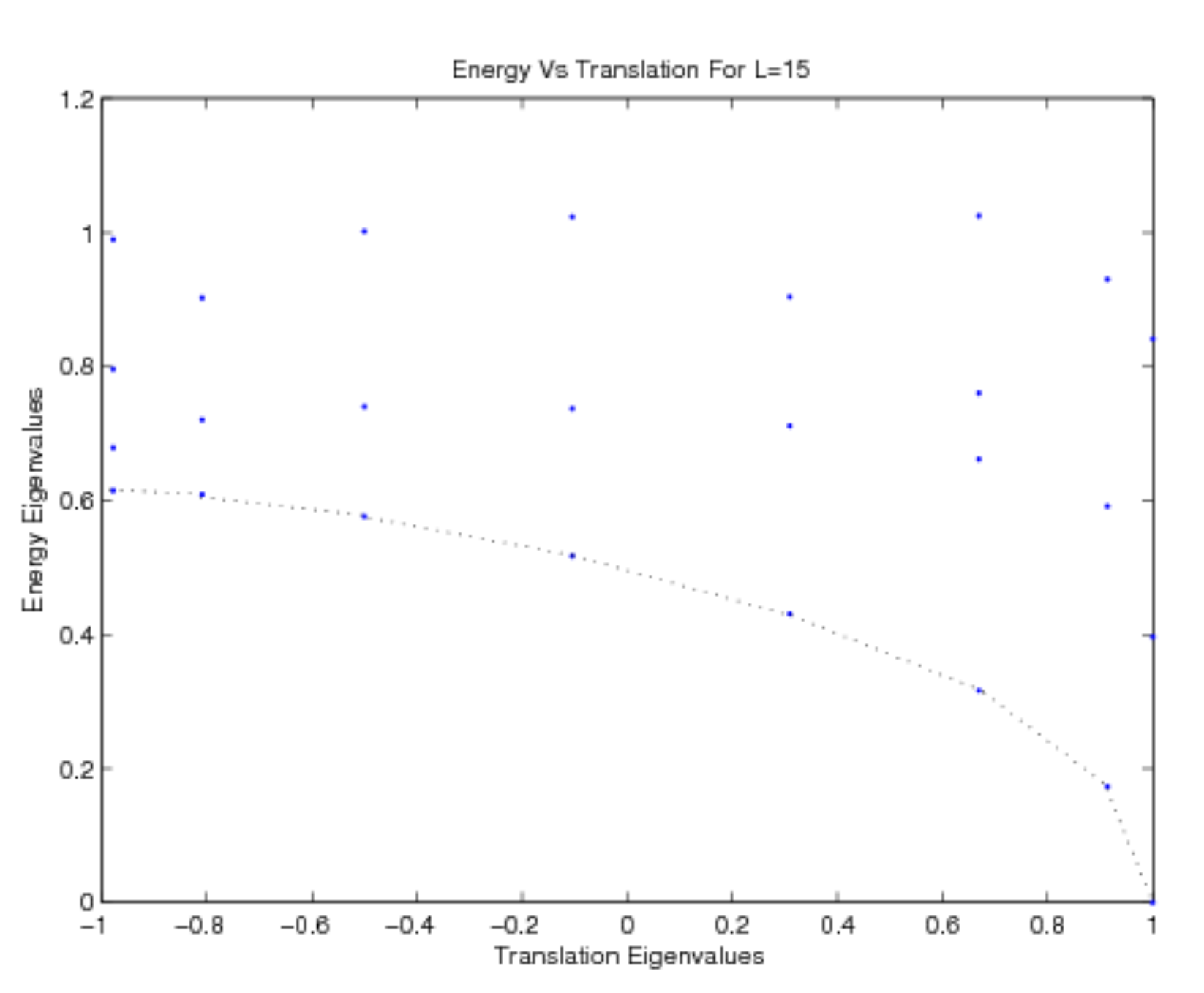}};
\end{tikzpicture}
\end{center}
\caption{\label{fig:third} We have plotted $\operatorname{spec}(2H \restriction \Hil_{\rm tot}(s,s))$ for the cycle $C_{15}$.
However, now the horizontal axis denotes $\cos \theta$ for the $T$ eigenvalue $e^{i\theta}$.
In this plot, we do not plot the total spin. But according to Sutherland's surmise, 
the minimum energy among all eigenvectors with $T \psi = e^{\pm 2\pi i k/15}$ is equal to the minimal
energy among all eigenvectors with total spin equal to $(15/2)-k$, for $k=0,1,\dots,7$.
Following this conjecture, we see that FOEL is satisfied for $N=15$.}
\end{figure}

In this section we report on results of numerical diagonalization of the spin-$1/2$ Heisenberg
model on cycles $C_N$ for various values of $N$, both even and odd.
We used Matlab's built-in ``eigs'' command which is an implementation of the Lanczos
iteration scheme. 
For $N=4,\dots,8$, we calculated the full spectrum.
This is plotted in Figure \ref{fig:first}.
For $N=10,12,14$, we have just plotted the lowest eigenvalues, and zoomed-in
to see the relative ordering of eigenvalues.
This is plotted in Figure \ref{fig:second}.
For $N=15$, we have plotted the eigenvalues separately, since size limitations required us to
plot only a subset of the relevant information.
This is shown in Figure \ref{fig:third}.
We now explain, point by point, the various features of relevance.

\subsection{Details of the implementation}

For $C_N$, we define $n=\lfloor N/2 \rfloor$.
When constructing the matrix for $H$ in Matlab, we restricted to the $n$-magnon subspace.
This is the subspace 
$$
\Hil_{\rm tot}^{\rm mag}((N/2)-n)\,
=\, \{\psi \in \Hil_{\rm tot}\, :\, S^z_{\rm tot} \psi = [(1/N)-n]\psi\}\, .
$$
Since total spin also commutes with $H$ and $S^z_{\rm tot}$, it is useful to consider the further
direct sum decomposition:
$$
\Hil_{\rm tot}^{\rm mag}((N/2)-n)\, =\, \bigoplus_{k=0}^{n} 
\Hil_{\rm tot}((N/2)-k,(N/2)-n)\, .
$$
Since $\operatorname{spec}(H \restriction \Hil_{\rm tot}(s,m)$
is actually independent of $m$, we see that in $\Hil_{\rm tot}^{\rm mag}((N/2)-n)$
we have representative eigenvectors corresponding to every eigenvalue of $H$
on the full Hilbert space $\Hil_{\rm tot}$.
So this is the optimal $S^z$ eigenspace to consider.

We used Matlab to do exact diagonalization. 
We also calculated the total spin of the eigenvectors.
Note that the translation operator $T$ also commutes with $H$, $S^{z}_{\rm tot}$ and ${\bf S}^2_{\rm tot}$.
So we also calculate the translation eigenvalues $e^{i\theta}$ for each eigenvector of $H$.
All of this is plotted in Figures 1 and 2.
On the horizontal axis we plotted total spin $s \in \{(N/2)-k\, :\, k \in \{0,1,\dots,n\}$,
such that for an eigenvector $\psi$ of $H$, we have
$$
{\bf S}^2_{\rm tot} \psi = s(s+1) \psi \, .
$$
On the vertical axis we have plotted the eigenvalues $E$ such that
$$
2 H \psi\, =\, E \psi\, .
$$
We plotted this in order to better compare to the $A$ eigenvalues from the last subsection.
It also matches the convention used by physicists, which we will review in the next section.

The plot points are open circles, and in each circle we have plotted rays associated to all the 
unit complex numbers $e^{i\theta}$ such that $T \psi = e^{i\theta} \psi$.
For $1$ and $-1$ the eigenvectors are typically non-degenerate.
But since $H$ is translation invariant and real, if $0<\theta<\pi$ and there is an eigenvector $\psi_+$
such that
$$
2H \psi_+\, =\, E \psi_+ \quad \text { and } \quad T \psi_{+}\, =\, e^{+i\theta} \psi_{+}\, ,
$$
then taking $\psi_-$ to be the complex conjugate of $\psi_+$, we have
$$
2H \psi_-\, =\, E \psi_- \quad \text { and } \quad T \psi_-\, =\, e^{-i\theta} \psi_-\, .
$$

For $N=15$, we were more limited in our use of Matlab.
Therefore, we did not plot the total spin.
We merely plotted the energy versus the translation eigenvalue.
In the next subsection we will explain why this also gives useful information.

\subsection{Some conjectures}

According to our numerical examples, we seem to see the following pattern:
\begin{itemize}
\item
For even length chains $C_{N}$ with $N=2n$,
$$
E_0(C_{2n},n)\, \leq\, E_0(C_{2n},n-1)\, .
$$
We found this to be true for $n=2,\dots,7$, and as we will describe, one may also infer
this violation of FOEL also for $n=8$.
\item
For odd length spin rings, $N=2n+1$, we did not find any violations of the FOEL ordering.
Specifically, we found $E_0(C_{2n+1},1/2) < E_0(C_{2n+1},3/2)$, which verifies FOEL.
This might simply be that we have not considered sufficiently long chains.
\end{itemize}

In order to explain the $N=15$ example, let us mention the following.
Physicists have considered a problem related to these counterexamples, although
not in the context of proving or disproving FOEL.
In particular, for the ring $C_N$, Sutherland \cite{Sutherland} noted the following trend.
Note that the $T$ eigenvalues are of the form $e^{i \theta}$ for $\theta = \pm 2 \pi k/N$
for $k=0,1,\dots,n$, where $n=\lfloor N/2 \rfloor$.
Let 
$$
\Hil_{\rm tot}^{\rm trans}(k)\, =\, \{ \psi \in \Hil_{\rm tot}\, :\, T \psi = e^{\pm 2 \pi i k/N} \psi\}\, ,
$$
for $k \in \{0,1,\dots,n\}$.
Then Sutherland's observation is
$$
\min \operatorname{spec}(H \restriction \Hil_{\rm tot}^{\rm trans}(k))\,
=\, \min \operatorname{spec}(H \restriction \Hil_{\rm tot}((N/2)-k,(N/2)-k))\, ,
$$
for $k \in \{0,1,\dots,n\}$.
To the best of our knowledge,  Sutherland did not prove this conjecture, and we also have not obtained
a rigorous proof.
But Sutherland's surmise is verified in all our examples $N=4,5,6,7,8,10,12,14$ as direct inspection
of Figure \ref{fig:first} and Figure \ref{fig:second} reveal.

In Figure \ref{fig:third} on the horizontal axis we plotted $\cos \theta$ for the $T$ eigenvalues $e^{i\theta}$.
Therefore, the lowest band of energy eigenvalues in Figure \ref{fig:third} should correspond to 
$E_0(C_{15},k)$ for $k=0,1,\dots,7$ reading from right-to-left (for increasing $k$ in $\cos(2\pi k/15)$).
Therefore, this does not appear to violate FOEL.

Finally, Dhar and Shastry reconsidered some of Sutherland's results in \cite{DharShastry}.
In their Figure 2, they considered the spin ring $C_{16}$, and plotted
energy versus translation eigenvalue.
One may clearly see from their figure that $E_0(C_{16},8) < E_0(C_{16},7)$, giving one more
example of violation of FOEL.
We did not attempt to reproduce their picture here.
Firstly, we were not able to numerically implement the ring for 16 sites.
But more importantly, the interested reader will do better to read their very beautiful paper,
which is available online. See reference \cite{DharShastry}.

\section{Relation to the ``spectral curve'' and violation of the string hypothesis in Bethe's ansatz}

These small calculations beg the question whether these violations of FOEL persist, so that there are always violations
of FOEL between levels $n-1$ and $n$ at $N=2n$ for $n>2$.
We believe that is true.
But they also beg the question whether this is a consequence of finite size corrections, or whether this has to do
with the limiting ``spectral'' curves.
This is supposedly answered by Sutherland \cite{Sutherland}, Dhar and Shastry \cite{DharShastry},
and Bargheer, Beisert and Gromov \cite{BargheerBeisertGromov}.

Sutherland deduced that for sufficiently large spin deviations, asymptotically, the density of the string
state would have to become greater than 1 in order to satisfy the string hypothesis.
But this is impossible because the roots must satisfy a trigonometric identity that has a minimum spacing of 1.
Therefore, Sutherland claims that there is a value of $\rho_* \in [0,1/2]$,
such that if one considers
$$
E_0(C_N,(N/2)-k)\, ,
$$
for $k\geq \rho_* N$, and sufficiently large $N$, then one needs to calculate a different ``spectral curve,''
than the vertical strings typically assumed to describe the Bethe roots in the so-called ``string hypothesis.''
Sutherland claims that the determination of the spectral curve is an example of a classical Cauchy problem,
and finds elliptic curves which give the correction to the string hypothesis.

In the last displayed equation of his paper he gives the following parametric formula for the density
of spin deviates and the energy density:
\begin{gather*}
d\, =\, (1/2) + a [ ( E(1/a) / K(1/a)) - 1]/2\, ,\\
\varepsilon\, =\, 4 K(1/a) [ 2 E(1/a) - (1 - (1/a^2))K(1/a)]\, .
\end{gather*}
Here the number of spin deviates is $n = dN$ and the energy is $E = \varepsilon/N$.
The energy scaling is reasonable because the spectral gap is $\gamma = O(1/N^2)$, and one expects the minimum
energy of singlets to be on the order of $O(\gamma N)$ because one adds an amount of energy
on the order of $O(\gamma)$ for each ``spin deviate''.
(See, for example, our forthcoming paper \cite{NSS3} for a rigorous derivation of some parts of the linear spin wave hypothesis
which suggests this energy scaling.)

Sutherland's formula involves $K(k)$ and $E(k)$, the complete elliptic integrals of first and second kind, respectively.
Sutherland notes that to obtain $d=1/2$ means one takes the parameter $a \to \infty$.
Physically, the parameter $a$ is related to the length of the spectral curve along which the Bethe roots accumulate.
At half-filling of spin deviates this is supposed to diverge.
But for small values of $k$, one has the asymptotic expressions:
\begin{equation}
\label{eq:Sutherland}
\begin{gathered}
K(k)\, =\, \frac{\pi}{2} \left(1 + \left(\frac{1}{2}\right)^2 k^2 + \left(\frac{1\cdot 3}{2\cdot 4}\right)^2 k^4 + \dots + \left(\frac{(2n-1)!!}{(2n)!!}\right)^2 k^{2n} + \dots \right)\, ,\\
E(k)\, =\, \frac{\pi}{2} \left(1 - \left(\frac{1}{2}\right)^2 \frac{k^2}{1} - \left(\frac{1\cdot 3}{2\cdot 4}\right)^2 \frac{k^4}{3} - \dots - \left(\frac{(2n-1)!!}{(2n)!!}\right)^2 \frac{k^{2n}}{2n-1} - \dots \right)\, .
\end{gathered} 
\end{equation}
Using this, one does recover $d=1/2$ as $a\to \infty$, which means $k=1/a \to 0$.
Also, one recovers Sutherland's formula that 
$$
\varepsilon \to \pi^2 \qquad \text { as }\quad  d \to 1/2\, ,
$$ 
in the large $N$ limit.

The curve above refers only to the minimum energy among eigenvectors with total angular momentum $\pi$.
According to Sutherland's surmise, this gives exactly the correct formula at half-filling of spin deviates, i.e., in the subspace with total spin equal to zero.
In particular, this means that $\varepsilon \to \pi^2$ as $d \to 1/2$ is correct within the context of the Bethe ansatz, using Sutherland's correction of the string
hypothesis.
Moreover, considering this curve for $d<1/2$ but very close to $1/2$, one sees that it is monotone decreasing.
On the other hand, the first derivative is zero.
Therefore, one needs to use a fine asymptotic analysis to see the persistence of the violation of the FOEL.

Dhar and Shastry continued Sutherland's work, starting from the ``low density'' regime.
They found numerically, that the value of $\rho_*$ where the string solution breaks down is around $1/4$.
At low densities, Dhar and Shastry discovered that the Bethe ansatz equations are asymptotically
satisfied by the Hermite points: roots of the Hermite functions. 
They also find that there is an exactly quadratic formula, as a function of the density of spin deviates,
at low density.
In fact, in Sutherland's notation, their formula is
$$
\varepsilon\, =\, 4 \pi^2 d(1-d)\, .
$$
Interestingly, their arguments indicate that this formula is not merely asymptotic in the limit $d\to 0$.
It should match with Sutherland's formula, if one includes the correct angular momentum curves
generalizing (\ref{eq:Sutherland}) for $d<1/2$.
(We are very grateful to an anonymous referee for clarifying this point to us, which we had mis-reported on in an earlier
version of this draft.)
To us, Dhar and Shastry's results seem to be wholly reliable.

Bargheer, Beisert and Gromov continued this line of investigation in \cite{BargheerBeisertGromov}.
They were motivated by new connections between models such as the spin-$1/2$ Heisenberg model and 
string theory apparently using the AdS/CFT correspondence.
But their analysis seems to be a very careful extension of Sutherland's orginal method.
In particular they give many details of the so-called ``spectral curves,'' that replace the vertical
lines corresponding to the string solution, at high density.
They also gave a concrete and simple numerical algorithm
for searching for the Bethe ansatz equations.
It amounts to fixing the spin deviate parameter $k$, and varying the length of the ring, $N$, treating it as a parameter.

For $N\gg k$, one is at ``low density.''
Therefore, Dhar and Shastry's Hermite solution is approximately valid.
This is the solution that Bargheer, Beisert and Gromov start with for a large value of $N$.
Then they run a Newton method solver to recover the exact formula.
At each step they decrease $N$, using the solutions of the Bethe ansatz equation
at the last step as the starting point for a Newton scheme.
As long as $N$ is sufficiently large, one may decrease $N$ in steps of size 1, proceeding along integers.
But as $N$ becomes smaller, this step size is too large to hope that the solution for $N+1$ is a good starting point for Newton's method for $N$.
Then they recommend to slowly decrease $N$, treating it as a real, continuous
parameter.
Of course for non-integer values of $N$, the Bethe ansatz equations are not algebraic.
But the Newton's method approach still seems to converge.
In our implementations of their method (both in Mathematica where they originally ran it, and in Matlab)
we found that the approach becomes chaotic as $N$ approaches $2k$ from above.
This seems to be part of the overall picture of a change of nature of this band of Bethe solutions
as $d$ approaches $1/2$.

The violations of FOEL for spin rings are tantalizingly close to these interesting results for the Bethe ansatz solution of the spin-$1/2$
Heisenberg model on rings.
We hope to return to this question later, or to discover a definitive answer within the physics literature.

\section*{Acknowledgements}

The research of 
S.S.\ was supported in part by a U.S.\ National Science Foundation
grant, DMS-0757327.
S.S.\ is grateful to S.G.~Rajeev for some useful conversations.
We are very grateful to an anonymous referee who corrected our earlier misinterpretation of 
reference \cite{Sutherland}.

\end{document}